\documentclass[aps,prb,twocolumn,groupedaddress,showpacs]{revtex4}
\usepackage{graphicx}
\usepackage{amsmath}
\newcommand{\comment}[1]{}
\begin{document}

\title{Two-channel Kondo physics in odd impurity chains}

\author{Andrew K. Mitchell,$^{1,2}$ David E. Logan$^1$ and H. R. Krishnamurthy$^3$}

\affiliation{$^1$Department of Chemistry, Physical and Theoretical
  Chemistry, Oxford University, South Parks Road, Oxford OX1 3QZ,
  United Kingdom\\
$^2$Institute for Theoretical Physics, University of Cologne, 50937
Cologne, Germany\\
$^3$Department of Physics, Indian Institue of Science, Bangalore 560 012, India}

\date{\today}


\begin{abstract}
We study odd-membered chains of spin-$\tfrac{1}{2}$ impurities, with each end connected to its own metallic lead. 
For antiferromagnetic exchange coupling, universal two-channel Kondo (2CK) physics is shown to arise at low
energies. Two overscreening mechanisms are found to occur depending on coupling strength, with distinct signatures in physical properties. For strong inter-impurity coupling, a residual chain spin-$\tfrac{1}{2}$ moment
experiences a renormalized effective coupling to the leads; while in the weak-coupling regime, Kondo coupling is mediated via incipient single-channel Kondo singlet formation. We also investigate models where the leads are tunnel-coupled to the impurity chain, permitting variable dot filling under applied gate voltages. Effective low-energy models for each regime of filling are derived, and for even-fillings where the chain ground state is 
a spin singlet, an orbital 2CK effect is found to be operative. Provided mirror symmetry is preserved, 2CK physics is shown to be wholly robust to variable dot filling; in particular the single-particle spectrum at the Fermi level, and hence the low-temperature zero-bias conductance, is always pinned to half-unitarity. We derive a 
Friedel-Luttinger sum rule and from it show that, in contrast to a Fermi liquid, the Luttinger integral
is non-zero and determined solely by the `excess' dot charge as controlled by gate voltage. The relevance of the work to real quantum dot devices, where inter-lead charge-transfer processes fatal to 2CK physics are present, is also discussed. Physical arguments and numerical renormalization group techniques are used to obtain a
detailed understanding of these problems.
\end{abstract}


\maketitle
\section{Introduction}\label{intro}
Non-Fermi liquid (NFL) behavior is famously realised in the two-channel Kondo (2CK) model,\cite{2ck:nozieres} in which a single spin-$\tfrac{1}{2}$ impurity is exchange-coupled to two equivalent but independent metallic conduction bands. Its fascination for theorists is evident in the wide range of techniques applied to 
the model, including notably the Bethe
ansatz,\cite{2ck:BA_sacramento,2ck:BA_andrei_jerez,2ck:BA_destri,2ck:BA_tsvel_wieg,2ck:BAnrg_zarand} 
numerical renormalization group\cite{2ck:nrg_cragg1,2ck:nrg_cragg2,2ck:nrg_pang_cox,2ck:BAnrg_zarand,2ck:CFT_NRG}
and conformal field theory\cite{2ck:CFT_NRG,2ck:CFT_affleck1,2ck:CFT_affleck3,2ck:CFT_affleck2} (for a 
review, see Ref.~\onlinecite{2ck:rev_cox_zaw}). Such methods have elucidated key aspects of the NFL
state arising from `overscreening' of the impurity spin below the low-energy 2CK scale,\cite{2ck:nozieres} $T_K^{2CK}$; including exotic physical properties such as a residual entropy of $\tfrac{1}{2}\ln(2)$, and a
logarithmically divergent low-temperature magnetic susceptibility, which are symptomatic of the frustration inherent when two conduction channels compete to screen the impurity local moment.\cite{2ck:rev_cox_zaw} NFL scaling of 
the conductance has also been predicted
theoretically\cite{2ck:CFT_affleck2,2ck:rev_cox_zaw,2ck:proposal,2ck:cond_pustilnik,2ck:dyn_toth},
and measured experimentally\cite{2ck:expt_potok} in quantum dots, with its square-root temperature dependence being a
characteristic signature of the 2CK phase.

This is all of course in marked contrast to standard Fermi liquid (FL) behavior,\cite{hewson} arising for example in the single-channel spin-$\tfrac{1}{2}$ Kondo or Anderson models\cite{hewson} (realised in
practice in e.g.\ ultrasmall quantum dots\cite{1dot:Goldhaber,1dot:Cronenwett}). Here, the dot spin is 
completely screened by a single bath of conduction electrons. The impurity entropy is in consequence quenched on the lowest energy scales, and the susceptibility is constant.\cite{hewson} Such systems are characterised by a unitarity zero-bias conductance, with a quadratic temperature dependence at low-energies.\cite{kondo:rev_pustilnik} 

But the NFL physics of the 2CK model is delicate: finite channel-asymmetry and/or inter-channel charge transfer ultimately drive any real system out of the NFL regime,\cite{2ck:rev_cox_zaw,2ck:CFT_affleck1} to a FL ground state.
The exquisite tunability of quantum dot devices\cite{rev:mesoscopic} allows manipulation of such perturbations; indeed couplings can be fine-tuned via application of gate voltages to effectively eliminate channel-asymmetry. 
However, tunnel-coupling in such systems must result in some degree of charge transfer between the metallic `leads'. 
This is of course responsible for the predominance of \emph{single}-channel Kondo physics in real quantum dot systems.

Suppressing inter-channel charge transfer allows for the emergence of 2CK physics at intermediate temperatures/energies, although the instability of the 2CK fixed point to charge-transfer means that an incipient NFL state forming at $T\sim T_K^{2CK}$ is subsequently destroyed below a FL crossover scale $T_{\text{FL}}$.  
Observation of NFL behavior at higher temperatures thus depends on a clear separation of
scales, $T_{\text{FL}}\ll T_K^{2CK}$. This was achieved recently~\cite{2ck:expt_potok} through use
of an \emph{interacting} lead tuned to the Coulomb blockade regime, and to date it is the only unambiguous experimental demonstration of the 2CK effect. Alternatively, sequential tunneling through several coupled dots should suppress charge transfer between leads,\cite{2ck:chains_zarand} and this could be exploited to access
2CK physics; although the interplay between spin and orbital degrees of freedom in coupled quantum dot systems can also generate new physics, as well known both
theoretically\cite{2IK:jones1,2IK:jones2,2ck:chains_zarand,DD:eran1,DD:eran2,vojta2spin,isingDD,galpinccdqd,akm_ccdqd,side:boese,side:cornaglia,side:zb,fanokondoDD,finitehub,ferrodots,zitkoTQD,akm_3dots,OguriTQD2009,TQD:FLNFL_zitko,zitkoTQD2ch,Martins2009,akm_frust,rev:nutshell} and
experimentally.\cite{blick2dot,A-B:LPK,artmol,vidan3dot,ludwig3dot,rogge3dot,nanotube:Grove,QPT:dots_nature,3dot:coher,3dot:granger,2ck:expt_potok}

In light of the above, we here consider odd-membered coupled quantum dot chains (each end of which is connected to its own metallic lead), and demonstrate that 2CK physics is indeed generally accessible in these systems.
In models where the couplings are of pure exchange type, we show that the low-energy behavior is described by the channel-asymmetric two-channel Kondo model,\cite{2ck:rev_cox_zaw} with pristine 2CK physics surviving down to the lowest energy scales in the mirror-symmetric systems of most interest. Such models are considered  in Sec.~\ref{chains}, where analytic predictions are confirmed and supplemented by use of Wilson's  numerical renormalization group (NRG) technique\cite{nrgreview,KWW,nrg_rev,asbasis,fdmnrg} 
(for a recent review see Ref.~\onlinecite{nrg_rev}). In particular, universal scaling of thermodynamic and dynamic properties is demonstrated for odd chains of different length, and for systems with different coupling strengths. 
While the underlying 2CK physics of such systems is shown to be robust for finite antiferromagnetic 
exchange coupling, we find that the mechanism of overscreening differs according to whether the inter-impurity coupling is strong or weak. In the former case (Sec.~\ref{largeJ}), it is the single lowest spin-$\tfrac{1}{2}$ state of the chain which effectively couples to and is overscreened by the leads;
while for weak inter-impurity coupling (Sec.~\ref{smallJ}), two-channel Kondo coupling is mediated via
incipient \emph{single}-channel Kondo singlets. Clear signatures of the latter are evident in the behavior of the frequency-dependent t-matrix, results for which are presented and analysed.

In order to investigate conductance across two-channel coupled quantum dot chains, we study in Sec.~\ref{GateV} a related class of models in which the terminal impurities are Anderson-like quantum \emph{levels}, tunnel-coupled to their respective metallic leads (although with  inter-lead charge transfer still precluded by inter-impurity exchange couplings). In the mirror-symmetric systems considered, we derive effective low-energy 2CK models valid for each regime of electron-filling. Even-occupation filling regimes -- where the
chain ground state is a spin-singlet -- are found in particular to exhibit an \emph{orbital}
2CK effect, with spin playing the role of a channel index. In consequence, 2CK physics is found to be robust throughout all regimes of electron-filling induced by changes in gate voltage.

Single-particle dynamics for such systems are then considered, and hence conductance (Sec.~\ref{GateV_Gc}), 
the S-matrix and associated phase shifts (Sec.~\ref{Smatrix}); again highlighting the universality arising at
low-energies. It is found in particular that the Fermi level value of the $T=0$ single-particle
spectrum -- and hence the zero-bias conductance -- is pinned to a half-unitary value, irrespective of 
electron-filling. A Friedel-Luttinger sum rule~\cite{CJW_multilevel}
is then derived in Sec.~\ref{GateV_lutt}, relating the Fermi level
value  of the spectrum to the `excess' charge due to the quantum dot
chain, and the Luttinger integral.\cite{lutt,lutt_ward} By virtue of
the spectral pinning, the sum rule relates directly the Luttinger
integral to the excess charge/dot filling; in contrast to a Fermi
liquid, where it is the Luttinger integral that is ubiquitously
`pinned' (to zero),\cite{lutt,lutt_ward} and the dot filling then
determines the value of the spectrum at the Fermi level.\cite{hewson,lang}

Finally, in Sec.~\ref{real dots} we consider briefly the applicability of our findings  to real coupled quantum dot devices. We argue that the effective low-energy model describing such systems is generically a
2CK model with both channel-asymmetry and inter-lead cotunneling charge transfer. Via a suitable basis transformation, one obtains a model in which charge transfer between even and odd channels is eliminated, whence the underlying behavior is readily understood in terms of that of a pure channel-asymmetric
2CK model. The spectrum/t-matrix in a given physical channel is however related via this transformation to a combination of t-matrices in even and odd channels. In the mirror-symmetric case sought experimentally, this leads to the striking conclusion that for sufficiently small but non-vanishing cotunneling charge transfer,
the crossover \emph{out} of the NFL regime is not in fact apparent in conductance measurements, despite the ultimate low-energy physics being that of a Fermi liquid.


\section{2CK Heisenberg chains}\label{chains}
\begin{figure}
\includegraphics[height=4cm]{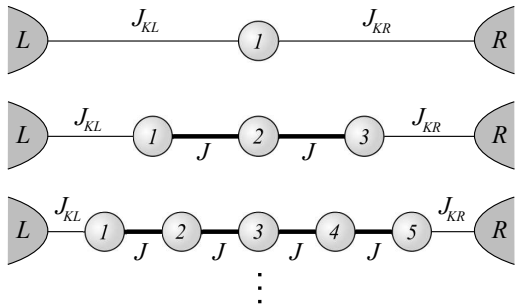}
\caption{\label{dots}
Schematic illustration of odd-membered impurity chains, each end of which is
coupled to its own metallic lead.
}\end{figure}

We consider a chain of $N_{c}$ coupled spin-$\tfrac{1}{2}$ impurities, each end of
which is also coupled to its own metallic lead, as illustrated in
Fig.~\ref{dots}. To investigate 2CK physics, we study explicitly in this section 
a system of exchange-coupled spin-$\tfrac{1}{2}$ impurities, where inter-lead charge transfer 
is eliminated from the outset. For an impurity chain of length $N_c$, the full Hamiltonian is thus  
$H^{N_c}=H_L+H^{N_c}_{\text{c}}+H^{N_c}_{\text{K}}$. Here
$H_L=\sum_{\alpha,\textbf{k},\sigma}\epsilon^{\phantom{\dagger}}_{\textbf{k}}
a^{\dagger}_{\alpha \textbf{k} \sigma}a^{\phantom{\dagger}}_{\alpha
  \textbf{k} \sigma}$ refers to the two equivalent non-interacting metallic
leads ($\alpha=L$/$R$), and
\begin{subequations}
\label{Hfull}
\begin{align}
H^{N_c}_{\text{c}} = & J \sum_{i=1}^{N_c-1}\hat{\textbf{S}}_{i} \cdot \hat{\textbf{S}}_{i+1}\\
H^{N_c}_{\text{K}} = & J_{KL} \hat{\textbf{S}}_{1} \cdot
\hat{\textbf{s}}_{L0} + J_{KR} \hat{\textbf{S}}_{N_c} \cdot
\hat{\textbf{s}}_{R0}
\end{align}
\end{subequations}
where $\hat{\textbf{S}}_i$ is a spin-$\tfrac{1}{2}$ operator for
impurity $i$, and $\hat{\textbf{s}}_{\alpha 0}$ is the spin density 
of lead $\alpha=L(R)$ at impurity $1(N_c)$:
\begin{subequations}
\label{0 orb}
\begin{align}
\label{0 orb S}
\hat{\textbf{s}}_{\alpha 0}&=\sum_{\sigma,\sigma'} f^{\dagger}_{\alpha 0 \sigma} \boldsymbol{\sigma}_{\sigma\sigma'} f^{\phantom{\dagger}}_{\alpha 0 \sigma'}\\
\label{0 orb f}
f^{\dagger}_{\alpha 0 \sigma}&=\tfrac{1}{\sqrt{N}}\sum_{\textbf{k}} a^{\dagger}_{\alpha\textbf{k}\sigma}~,
\end{align}
\end{subequations}
with $\boldsymbol{\sigma}$ the Pauli matrices and $f^{\dagger}_{\alpha
  0 \sigma}$ the creation operator for the `0'-orbital of the
$\alpha=L$/$R$ Wilson chain ($N\rightarrow\infty$ is the number of orbitals/$\mathbf{k}$-states in the chain).

In the following, we focus on odd-$N_c$ chains, with
antiferromagnetic (AF) exchange couplings for both the intra-chain
Heisenberg exchange, $J>0$, and the Kondo exchanges,
$J_{K\alpha}>0$. [We do not consider here even-$N_c$ chains, where the
generic low-energy physics is quite different and in essence that
of the two-impurity Kondo model]. We also consider both the channel
symmetric case, $J_{KR}=J_{KL}$, as well as the more general case
where channel asymmetry is present, $J_{KR}\ne J_{KL}$. The
simplest member of the family, $N_c=1$, is of course the classic
single-spin 2CK model,\cite{2ck:nozieres,2ck:rev_cox_zaw}
while variants of the $N_c=3$ trimer have also been considered
previously.\cite{ferrodots,zitkoTQD,akm_3dots,OguriTQD2009,TQD:FLNFL_zitko,zitkoTQD2ch,Martins2009,akm_frust} 
We show below that an effective single-spin 2CK model results in all
cases, whence universal 2CK physics is expected on the lowest
energy scales in the channel-symmetric case; albeit that 
the mechanism by which the effective two-channel coupling arises 
is rather different in the strong and weak inter-impurity coupling regimes.

\subsection{Strong inter-impurity coupling}\label{largeJ}
We consider first the case where the inter-impurity exchange couplings $J$ are 
sufficiently large that only the ground state of the isolated spin chain is relevant in
constructing the effective low-energy model upon coupling to the
leads. As detailed in Sec.~\ref{smallJ}, this means in practice $J
\gtrsim T_{K,\alpha}^{1CK}$, with $T_{K,\alpha}^{1CK}$ the scale for
single-channel Kondo quenching of a terminal spin to lead $\alpha$, 
arising in the ``uncoupled'' $J=0$ limit. The lowest 
state of an AF-coupled odd-membered spin-$\tfrac{1}{2}$ chain is of course a spin doublet,
the components of which we label as $|N_c;S^z=\pm\tfrac{1}{2}\rangle$. 
All other states are at least $\mathcal{O}(J/N_c)$ higher in energy.


\subsubsection{Effective 2CK model}\label{largeJ eff}
To leading order in $J_{K\alpha}$, the low-energy model is
then obtained simply by projecting into the reduced Hilbert space of the
lowest doublet for a chain of length $N_c$, using
\begin{equation}
\label{unity}
\hat{1}_{N_c}=\sum_{S^z} |N_c;S^z\rangle \langle N_c;S^z|.
\end{equation}
The resultant Hamiltonian $H^{N_c}_{\text{eff}}=\hat{1}_{N_c} H^{N_c}_{c} \hat{1}_{N_c}$ follows as
\begin{equation}
\label{Heff unity}
\begin{split}
H^{N_c}_{\text{eff}}=\tfrac{1}{2}J_K
\hat{1}_{N_c}&(\hat{\textbf{S}}_{1}+\hat{\textbf{S}}_{N_c})\hat{1}_{N_c}\cdot(\hat{\textbf{s}}_{L0}+
\hat{\textbf{s}}_{R0})\\
+ \tfrac{1}{2}\delta_K \hat{1}_{N_c}&(\hat{\textbf{S}}_{1}+\hat{\textbf{S}}_{N_c})\hat{1}_{N_c}\cdot(\hat{\textbf{s}}_{L0}
- \hat{\textbf{s}}_{R0})
\end{split}
\end{equation}
where $J_K=\tfrac{1}{2}(J_{KL}+J_{KR})$ and
$\delta_K=\tfrac{1}{2}(J_{KL}-J_{KR})$, and we use
the symmetry  $\hat{1}_{N_c}\hat{\textbf{S}}_{1}\hat{1}_{N_c}=
\hat{1}_{N_c}\hat{\textbf{S}}_{N_c}\hat{1}_{N_c}$. 

In the absence of a magnetic field, $\uparrow/\downarrow$ 
spin symmetry implies $\hat{P}_{\uparrow\downarrow}|N_c;S^z\rangle=\gamma 
|N_c;-S^z\rangle$, where $\hat{P}_{\uparrow\downarrow}$
permutes simultaneously all up and down spins, and
$\gamma=\pm 1$ only since $\hat{P}_{\uparrow\downarrow}^2=\hat{1}$.\cite{akm_frust} Together
with $\hat{P}_{\uparrow\downarrow}\hat{S}^z_i=-\hat{S}^z_i\hat{P}_{\uparrow\downarrow}$, it
follows directly that
\begin{equation}
\begin{split}
\label{matrix elem}
\langle N_c;S^z&| \hat{S}_1^z+\hat{S}_{N_c}^z |N_c ; S^z \rangle \\
= - \langle N_c;-S^z&| \hat{S}_1^z+\hat{S}_{N_c}^z |N_c ; -S^z \rangle \propto S^z
\end{split}
\end{equation}
(as $|N_{c};S^{z}\rangle$ is a spin doublet).
Such matrix elements appear in the $z$-component of the contraction in
Eq.~\ref{Heff unity}, and by spin isotropy an effective model of 2CK form results:
\begin{equation}
\label{Heff bigj}
H^{N_c}_{\text{eff}}=J_{K,N_c}^{\text{eff}} \hat{\textbf{S}}\cdot(\hat{\textbf{s}}_{L0}+
\hat{\textbf{s}}_{R0}) + \delta^{\text{eff}}_{K,N_c} \hat{\textbf{S}}\cdot(\hat{\textbf{s}}_{L0}
- \hat{\textbf{s}}_{R0})
\end{equation}
where $\hat{\textbf{S}}$ is a spin-$\tfrac{1}{2}$ operator for the
lowest chain doublet, defined by $\hat{S}^z = \sum_{S^{z}}|N_c;S^z\rangle
S^z \langle N_c;S^z|$ and $\hat{S}^{\pm} = |N_c;\pm\tfrac{1}{2}\rangle
\langle N_c;\mp\tfrac{1}{2}|$. The effective couplings are given by
\begin{subequations}
\label{eff coup bigj}
\begin{align}
J_{K,N_c}^{\text{eff}}=\langle N_c;+\tfrac{1}{2}&| \hat{S}_1^z+\hat{S}_{N_c}^z |N_c ; +\tfrac{1}{2} \rangle J_K,\\
\delta_{K,N_c}^{\text{eff}}= \langle N_c; +\tfrac{1}{2}&| \hat{S}_1^z+\hat{S}_{N_c}^z |N_c ;
+\tfrac{1}{2}\rangle \delta_K.
\end{align}
\end{subequations}
Numerical evaluation of Eq.~\ref{eff coup bigj} for odd $N_c$ 
yields an AF effective coupling, $J_{K,N_c}^{\text{eff}}>0$, which
is renormalized with respect to the bare coupling and
diminishes as the chain length increases, as shown in Table I. 
\begin{table}[h]
\caption{Effective couplings}
\centering
\begin{tabular}{l l}
\hline\hline 
$N_c$ & \qquad $J_{K,N_c}^{\text{eff}}/J_{K}^{\phantom\dagger}$ \\ [0.5ex] 
\hline 
1 & \qquad $1$ \\
3 & \qquad $\tfrac{2}{3}$ \\
5 & \qquad $0.51...$ \\
7 & \qquad $0.42...$ \\ [1ex] 
\hline 
\end{tabular}
\label{table:coup}
\end{table}

Hence, for sufficiently low temperatures $T\ll J/N_c$, the low-energy
behavior of the system is equivalent to the single-spin 2CK model.\cite{2ck:rev_cox_zaw}
In the $L/R$ mirror-symmetric case, $J_{KL}=J_{KR}$ 
($\delta_K=\delta_{K}^{\text{eff}}=0$), the stable $T=0$ fixed point 
(FP) is then the infrared 2CK FP. The lowest spin-$\tfrac{1}{2}$ state of the
impurity chain is thus overscreened by conduction electrons; giving rise to a residual entropy of
$S_{\text{imp}}=\tfrac{1}{2}\text{ln}(2)~~(k_B=1)$, a hallmark of the NFL 2CK ground
state.\cite{2ck:rev_cox_zaw} Overscreening sets in below a characteristic scale $T_K^{2CK}$, 
given from perturbative scaling\cite{2ck:nozieres} as
\begin{equation}
\label{2CK pms}
T_K^{2CK}\sim J_{K,N_c}^{\text{eff}} \exp(-1/\rho J_{K,N_c}^{\text{eff}} ),
\end{equation}
where $\rho=1/(2D)$ is the lead density of states per orbital 
(assumed to be uniform) and $2D$ is the bandwidth.

By contrast, when strict $L/R$ mirror symmetry is broken via distinct
exchange couplings to the two leads, $J_{KL}\ne J_{KR}$ (ie.\ $\delta_K \ne 0$),
the 2CK FP is destabilized in the full model Eq.~\ref{Hfull}. This behavior
is of course well-known from the single-spin 2CK model with channel anisotropy,\cite{2ck:nozieres,2ck:rev_cox_zaw,2ck:BA_sacramento,2ck:BA_andrei_jerez,2ck:CFT_NRG} where the impurity local moment is fully screened by conduction electrons in the more strongly coupled lead, and a Fermi liquid ground state results. For $\delta_K>0$, under renormalization on reduction of the
temperature/energy scale, the system flows to \emph{strong coupling}
(SC) with the left lead ($J_{KL}\rightarrow \infty$) while the right lead
decouples ($J_{KR}\rightarrow 0$).\cite{2ck:nozieres,2ck:rev_cox_zaw,2ck:BA_sacramento,2ck:BA_andrei_jerez,2ck:CFT_NRG} The stable $T=0$ FP thus depends on the \emph{sign} of $\delta_K$, with `SC:L' describing the lowest energy
behavior for $\delta_K>0$ while `SC:R' is stable for
$\delta_K<0$. The mirror-symmetric case, $\delta_K=0$, is as such the
quantum critical point separating phases where a Kondo singlet forms
in either the $L$ or $R$ lead.

In the full model, effective \emph{single-channel} Kondo screening characteristic of flow to the Fermi liquid FP in channel-asymmetric systems, sets in below a characteristic scale which can likewise be obtained from perturbative
scaling:\cite{2ck:nozieres}
\begin{equation}
\label{SC asym pms}
T_K^{SC}\sim J_{K,N_c}^{>} \exp(-1/\rho J_{K,N_c}^{>} ),
\end{equation}
where
$J_{K,N_c}^{>}=J_{K,N_c}^{\text{eff}}+|\delta_{K,N_c}^{\text{eff}}|$
is the effective coupling between the lowest chain doublet state and
the more strongly-coupled lead.


\begin{figure}
\includegraphics[height=5cm]{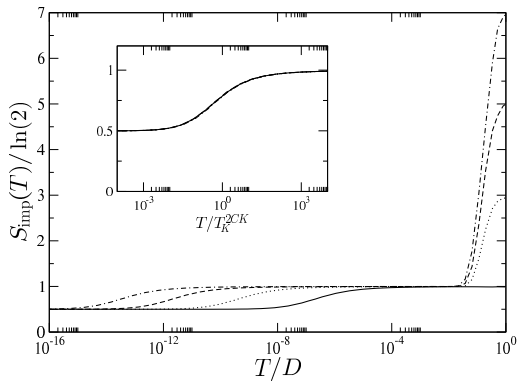}
\caption{\label{bigj ent fig}
Entropy $S_{\text{imp}}(T)/\ln(2)$ \emph{vs} $T/D$ for chains of length
$N_c=1,3,5,7$ (solid, dotted, dashed and dot-dashed lines respectively) in the
large inter-impurity coupling regime. Shown for
$\rho J=0.15$ and $\rho J_K = 0.075$ in the $L/R$-symmetric limit
$\delta_K=0$. Inset: scaling collapse onto the universal 2CK curve. 
}\end{figure}

\subsubsection{NRG results: symmetric case}\label{largeJ symm}
The above picture indicates that in the mirror symmetric
case, $\delta_K=0$, the lowest energy behavior for all odd chains
should be that of the single-spin 2CK model,\cite{2ck:rev_cox_zaw} but
with renormalized effective couplings (Eq.~\ref{eff coup bigj} and 
Table~\ref{table:coup}) and hence from Eq.~\ref{2CK pms} a reduced 
2CK scale, $T_K^{2CK}$.  

We now analyze the full model, Eq.~\ref{Hfull}, for odd $N_c=1,3,5,7$ using 
Wilson's  NRG technique,\cite{nrgreview,KWW,nrg_rev} employing a complete 
basis set of the Wilson chain\cite{asbasis} to
calculate the full density matrix.\cite{asbasis,fdmnrg}
Calculations here are typically performed using an NRG discretization 
parameter $\Lambda=3$, retaining the lowest $N_s=4000$ states per iteration. 
We consider first the impurity contribution\cite{KWW,nrg_rev} to thermodynamics
$\langle\hat{\Omega}\rangle_{\mathrm{imp}} =\langle\hat{\Omega}\rangle
- \langle\hat{\Omega}\rangle_{0}$, with
$\langle\hat{\Omega}\rangle_{0}$ denoting a thermal average in the
absence of the impurity chain. We focus in particular on the entropy,
$S_{\text{imp}}(T)$, and the uniform spin susceptibility, $\chi_{\text{imp}}(T)
=\langle (\hat{S}^{z})^{2}\rangle_{\mathrm{imp}}/T$ (here
$\hat{S}^{z}$ refers to the spin of the entire system); the
temperature dependences of which provide clear signatures of the
underlying FPs reached under renormalization on  progressive reduction
of the temperature/energy scale.\cite{KWW,nrg_rev}  

Fig.~\ref{bigj ent fig} shows representative NRG results for
the $T$-dependence of the entropy, for odd chains with $N_c=1,3,5,7$.
At high temperatures the impurities are effectively
uncoupled, so the chain contribution to the entropy is
$S_{\text{imp}}=N_c\ln(2)$, as seen clearly from Fig.~\ref{bigj ent
 fig}. On the scale of $T\sim J/N_c$, all but the lowest chain
doublet are projected out, the entropy then dropping as expected to
$\ln(2)$ in all cases. Renormalization group (RG) flow to this local
moment (LM) FP\cite{2ck:nozieres,KWW} marks the regime of validity of the effective
single-spin 2CK model, Eq.~\ref{Heff bigj}. The local moment is then
overscreened\cite{2ck:nozieres,2ck:rev_cox_zaw} by symmetric 
coupling to the two leads on an exponentially-small scale, $T_K^{2CK}$; the 
$T=0$ residual entropy in all cases being $\tfrac{1}{2}\ln(2)$. 
$T_K^{2CK}$ itself is determined in practice from $S_{\text{imp}}(T_K^{2CK})=\tfrac{3}{4}\ln(2)$
(suitably between the characteristic LM and 2CK FP values). The 
exponential reduction of the 2CK scale with increasing chain length (evident in 
Fig.~\ref{bigj ent fig}) is expected from
Eq.~\ref{2CK pms}, which depends on $N_c$ through the effective
coupling given in Table~\ref{table:coup}. The inset shows the data rescaled in
terms of $T/T_K^{2CK}$.
The low-temperature behavior of all odd chains collapse to the universal scaling curve
for the single-spin 2CK model (ie.\ the $N_c=1$ case, solid line), likewise known
from e.g.\ the Bethe ansatz
solution,\cite{2ck:BA_sacramento,2ck:BA_andrei_jerez,2ck:BA_destri,2ck:BA_tsvel_wieg,2ck:BAnrg_zarand}
confirming the mapping of the full model to Eq.~\ref{Heff bigj}.

\begin{figure}
\includegraphics[height=5cm]{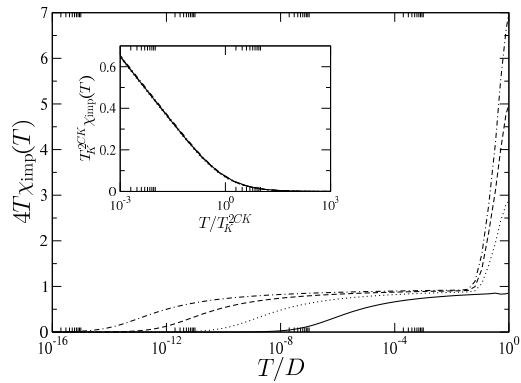}
\caption{\label{bigj susc fig}
Spin susceptibility $4T\chi_{\text{imp}}(T)$ \emph{vs} $T/D$ for the same
parameters as Fig.~\ref{bigj ent fig}. Inset: scaling collapse of
$\chi_{\text{imp}}(T)$ itself onto
the universal 2CK curve, with characteristic NFL form 
$T_K^{2CK}\chi_{\text{imp}}(T)=\alpha\ln(T_K^{2CK}/T)$.
}\end{figure}

The above FP structure and energy scales also naturally show up in the
magnetic susceptibility, as seen in Fig.~\ref{bigj susc fig}. At the
highest temperatures, $T\chi_{\text{imp}}=\tfrac{1}{4}N_c$, as
expected\cite{hewson,KWW} for $N_c$ free spins. Flow to the LM FP on the scale $T\sim J/N_c$
is again clearly evident in the drop to
$T\chi_{\text{imp}}=\tfrac{1}{4}$, corresponding to the expected Curie law
behavior. On the scale $\sim T_K^{2CK}$, the susceptibility  drops to
$T\chi_{\text{imp}}=0$, which remains its $T=0$ value. The inset
shows $T_K^{2CK}\chi_{\text{imp}}$ \emph{vs} $T/T_K^{2CK}$, showing scaling
collapse to the universal single-spin 2CK curve, and
demonstrating the characteristic NFL logarithmic divergence\cite{2ck:rev_cox_zaw} of the
susceptibility $\chi_{\text{imp}}$ itself on approaching the 2CK FP. \\

We turn now to dynamics, in particular the low-energy Kondo resonance embodied in the
spectrum $D\rho_K(\omega) \equiv D\rho_{K,\alpha}(\omega) =
- \pi\rho_T \text{Im}[t_{\alpha}(\omega)]$; where
$t_{\alpha}(\omega)\equiv t(\omega)$ is the t-matrix\cite{hewson} for the
$\alpha=L/R$ lead (equivalent by symmetry for $\delta_K=0$),
and $\rho_T=N\rho$ is the total lead density of states. Using equation of 
motion techniques\cite{hewson,EOM} yields
\begin{equation}
\label{t-matrix}
\pi\rho_T t_{\alpha}(\omega) = \frac{\pi}{4} \rho J_{K\alpha}^2
\tilde{G}_{\alpha i}(\omega)
\end{equation}
with $i=1$ for $\alpha=L$ and $i=N_c$ for $\alpha=R$, where
\begin{equation}
\label{Gi def}
\tilde{G}_{\alpha i}(\omega) = \langle\langle
\hat{S}_i^{-} f^{\phantom{\dagger}}_{\alpha 0\downarrow}+\hat{S}_i^z
f^{\phantom{\dagger}}_{\alpha 0\uparrow}; \hat{S}_i^{+}
f^{\dagger}_{\alpha 0\downarrow}+\hat{S}_i^z f^{\dagger}_{\alpha 0\uparrow} \rangle\rangle_{\omega}^{\phantom\dagger}
\end{equation}
and $\langle\langle \hat{A};\hat{B}
\rangle\rangle_{\omega}^{\phantom\dagger}$ is the Fourier transform of
the retarded correlator $\langle\langle \hat{A}(t_1);\hat{B}(t_2)
\rangle\rangle=-i\theta(t_1-t_2)\langle\{\hat{A}(t_1),\hat{B}(t_2)\}\rangle$.
The correlator in Eq.~\ref{Gi def} can be calculated directly within
NRG,\cite{asbasis,fdmnrg,nrg_rev} hence enabling access to the spectrum $D\rho_K(\omega)$.
However, an alternative expression for the t-matrix can be obtained in the spirit
of Ref.~\onlinecite{UFG}, and in the wide flat-band case considered here is simply
\begin{equation}
\label{t-matrix2}
\pi\rho_T t_{\alpha}(\omega) =
-i\left [1+\left ( \frac{2}{\pi\rho J_{K\alpha}} \right )^2 \frac{G_{\alpha
  0}(\omega)}{\tilde{G}_{\alpha i}(\omega)} \right ]^{-1}
\end{equation}
where $G_{\alpha 0}(\omega) = \langle\langle
f^{\phantom{\dagger}}_{\alpha 0\sigma};f^{\dagger}_{\alpha 0\sigma}
\rangle\rangle_{\omega}^{\phantom\dagger}$ is the Green function for
the `0'-orbital of the $\alpha=L/R$ Wilson chain.\cite{nrgreview} The
quotient of correlators in Eq.~\ref{t-matrix2} is found to improve 
greatly numerical accuracy, and is employed in the following.

\begin{figure}
\includegraphics[height=5cm]{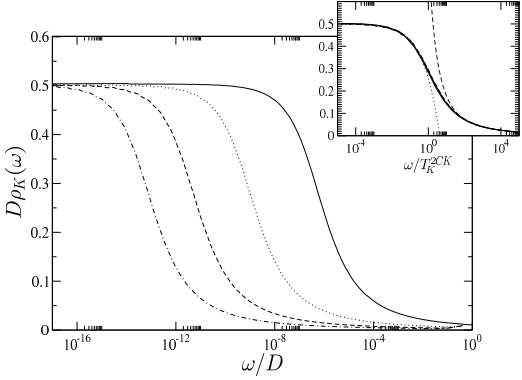}
\caption{\label{bigj spec fig}
$T=0$ spectrum $D\rho_K(\omega)$ \emph{vs} $\omega/D$, for the same parameters as
Fig.~\ref{bigj ent fig}. Inset: \emph{vs}  $\omega/T_K^{2CK}$, showing  collapse to the scaling spectrum for odd
chains. Grey dotted line:  low-$|\omega|/T_K^{2CK}\ll 1$
asymptotic behavior
$D\rho_K(\omega)=\tfrac{1}{2}[1-b(|\omega|/T_K^{2CK})^{1/2}]$; 
grey dashed line: high-$|\omega|/T_K^{2CK}\gg 1$ scaling
behavior, $D\rho_K(\omega)=A/[\ln^2(|\omega|/T_K^{2CK}) +B]$.
}\end{figure}

Fig.~\ref{bigj spec fig} shows the resultant spectrum $D\rho_K(\omega)$
\emph{vs} $\omega/D$ for chains of length $N_c=1,3,5,7$ with the same parameters as 
Figs.~\ref{bigj ent fig} and \ref{bigj susc fig} (noting that $\rho_K(\omega)=\rho_K(-\omega)$ since 
the model, Eq.~\ref{Hfull}, is particle-hole symmetric).
The low-energy form of each spectrum naturally reflects RG flow in the vicinity of the 2CK
FP, as studied also in a variety of different models which exhibit 2CK 
behavior.\cite{2ck:CFT_affleck2,TQD:FLNFL_zitko,2ck:dyn_johannesson,2ck:dyn_bradley,2ck:toth,akm_frust,2ck:dyn_anders}. In particular, for all odd chains, a half-unitarity value is seen to arise at the Fermi
level, $D\rho_K(\omega=0)=\tfrac{1}{2}$; and collapse to the universal single-spin 2CK scaling spectrum 
is clearly evident in the inset to Fig.~\ref{bigj spec fig}. The leading low-frequency
asymptotic behavior is 
$D\rho_K(\omega)=\tfrac{1}{2}[1-b(|\omega|/T_K^{2CK})^{1/2}]$ (in
marked contrast to the form $[1-a(|\omega|/T_K)^2]$ characteristic\cite{hewson} of
RG flow near a  Fermi liquid FP); and with which the numerics agree well for $\omega\ll T_K^{2CK}$.
At high frequencies $\omega\gg T_K^{2CK}$ by contrast, the leading asymptotic behavior of the scaling spectrum is
$D\rho_K(\omega)=A/[\ln^2(|\omega|/T_K^{2CK})+B]$, which behavior is common to other models in which spin-flip scattering processes are important at high energies,\cite{NLDscaling,LMA:SU(2N)}
such as the single-channel Anderson or Kondo models.

\begin{figure}
\includegraphics[height=5cm]{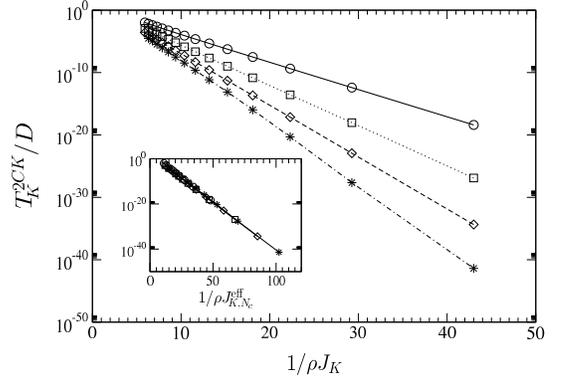}
\caption{\label{bigj tk fig}
Kondo temperature 
$T_K^{2CK}/D$ \emph{vs} $1/\rho J_K$ for chains of length
$N_c=1,3,5,7$ (circles, squares, diamonds and stars respectively, with
lines as guide to the eye) and common exchange coupling
$\rho J = 0.15$. Inset: scaling collapse to common linear form when plotted \emph{vs}
$1/\rho J_{K,N_c}^{\text{eff}}$ (given by Eq.~\ref{eff coup bigj}).
}\end{figure}

Finally, we consider the evolution of the 2CK scale itself as the
impurity-lead coupling is varied in the mirror-symmetric case, 
shown \emph{vs} $1/\rho J_K$ in Fig.~\ref{bigj tk fig} for chains of length $N_c=1,3,5,7$.
An exponential dependence of the 2CK scale on the impurity-lead coupling is 
expected from Eq.~\ref{2CK pms}, and seen clearly in the main panel. The differing slopes reflect the
renormalization of the bare Kondo coupling
$J_K\rightarrow J_{K,N_c}^{\text{eff}}$ with increasing impurity chain length 
 (Table~\ref{table:coup}); collapse to common linear behavior being observed in the inset 
where the 2CK scales are plotted \emph{vs} $1/\rho J_{K,N_c}^{\text{eff}}$, establishing
thereby quantitative agreement with Eqs.~\ref{eff coup bigj},\ref{2CK pms} and
Table~\ref{table:coup}, and hence the mapping to the effective 2CK model, Eq.~\ref{Heff bigj}.


\subsubsection{NRG results: asymmetric case}\label{largeJ asymm}
\begin{figure}
\includegraphics[height=6cm]{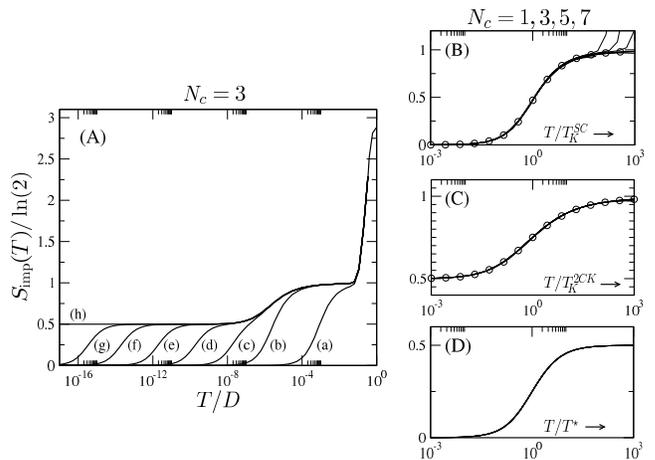}
\caption{\label{bigj asym ent fig}
(A): Entropy $S_{\text{imp}}(T)/\ln(2)$ \emph{vs} $T/D$ on progressively approaching
the transition, for $N_c=3$. Shown for fixed $\rho J=0.25$ and $\rho
J_K=0.125$, varying $\rho \delta_K=10^{-1}$, $10^{-2}$, $10^{-3}$,
$10^{-4}$, $10^{-5}$, $10^{-6}$, $10^{-7}$  [lines (a)--(g)], 
with the symmetric point $\rho \delta_K=0$ shown as line (h).
(B): Scaling collapse to the 1CK curve (circles) for strong channel asymmetry,
$\rho\delta_K=10^{-1}$ and $10^{-2}$, with $N_c=1,3,5,7$. 
(C): For small asymmetry ($\rho\delta_K=10^{-6}$ and $10^{-7}$), and
$N_c=1,3,5,7$. Showing universality in the LM$\rightarrow$2CK FP crossover,
on rescaling in terms of $T/T_K^{2CK}$; compared directly to the symmetric 2CK scaling curve
(circles).
(D): Same data as (C), now rescaled in terms of $T/T^{\star}$;  showing universal crossover from the 2CK FP 
($S_{\text{imp}}/\ln2 =\tfrac{1}{2}$) to the stable SC:L FP ($S_{\text{imp}}=0$).
}\end{figure}

We turn now to the channel-asymmetric case, $J_{KL}\ne J_{KR}$
(ie. $\delta_K\ne 0$). The effective model Eq.~\ref{Heff bigj}
should describe the low-energy behavior of all odd chains, so the rich
physics of the \emph{asymmetric} single-spin 2CK
model\cite{2ck:nozieres,2ck:rev_cox_zaw,2ck:BA_sacramento,2ck:BA_andrei_jerez,2ck:CFT_NRG}
is thus expected for $T\ll J/N_c$. As discussed above, breaking $L/R$ mirror
symmetry is a relevant perturbation\cite{2ck:CFT_NRG} to the 2CK
FP, so FL physics will arise generically on the lowest energy
scales. Indeed, in the limit of maximal asymmetry $J_K=\delta_K$, the
right lead is completely decoupled in Eq.~\ref{Heff bigj}; pristine
single-channel Kondo (1CK) screening by the left lead then results
below a single-channel scale $T_K^{1CK}$. For $\delta_K\ll T_K^{1CK}$, however, RG flow in
the vicinity of the 2CK FP strongly affects the behavior at higher
temperatures/energies.\cite{2ck:nozieres,2ck:rev_cox_zaw,2ck:BA_sacramento,2ck:BA_andrei_jerez,2ck:CFT_NRG}
This is indeed seen in Fig.~\ref{bigj asym ent fig}(A), where the 
entropy $S_{\text{imp}}(T)$ \emph{vs} $T/D$ is shown
for a representative system with $N_c=3$, $\rho J =0.25$, $\rho
J_K=0.125$, varying $\rho \delta_K = 10^{-1}$, $10^{-2}$, $10^{-3}$,
$10^{-4}$, $10^{-5}$, $10^{-6}$, $10^{-7}$  [lines (a)--(g)],
successively approaching the quantum critical point at the symmetric
limit, $\delta_K=0$: line (h).  
The scale $T_K^{1CK}$ (for which $\rho J_{K}=\rho\delta_{K}$) is $\rho T_K^{1CK} \approx 10^{-3}$ here;
and sets the scale for the crossover from `large' to `small' channel asymmetry ($\rho \delta_{K}\gg \rho T_K^{1CK}$
and $\ll \rho T_K^{1CK}$ respectively).

At the highest temperatures the three impurity spins are effectively
free, yielding trivially a common entropy $S_{\text{imp}}=3\ln(2)$. As $T$ is lowered, all but the
lowest trimer doublet state is projected out, heralding flow to the LM
FP, with characteristic\cite{2ck:nozieres,KWW} entropy $S_{\text{imp}}=\ln(2)$. For large
channel asymmetry [e.g.\ lines (a),(b)], RG flow is then directly to the SC:L FP:
the impurity spin is fully screened\cite{2ck:nozieres,2ck:rev_cox_zaw,2ck:BA_sacramento,2ck:BA_andrei_jerez,2ck:CFT_NRG} by formation of a Kondo
singlet with conduction electrons in the left lead ($\delta_K>0$) below an
effective single-channel Kondo scale $T_K^{SC}$, and hence
$S_{\text{imp}}=0$ for $T\ll T_K^{SC}$. The Kondo scale itself is
given by Eq.~\ref{SC asym pms} in the large $\delta_K$ regime,\cite{note:tkdef} with an 
effective Kondo coupling
$J_{K,N_c=3}^{\text{eff}}=\tfrac{2}{3}(J_K+|\delta_K|)$ for $N_c=3$
(see Table~\ref{table:coup}).

\begin{figure}
[ht]
\includegraphics[height=6cm]{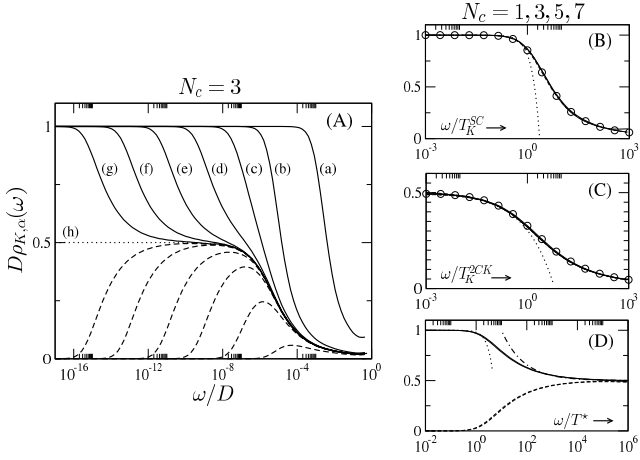}
\caption{\label{asym bigj spec}
(A): Spectra $D\rho_{K,L}(\omega)$ (solid lines) and
$D\rho_{K,R}(\omega)$ (dashed) \emph{vs} $\omega/D$ on progressively approaching
the transition for $N_c=3$, using the same parameters as in 
Fig.~\ref{bigj asym ent fig}. 
(B): Collapse of $D\rho_{K,L}(\omega)$ to the 1CK scaling spectrum (circles) in the strongly asymmetric limit,
$\rho\delta_K=10^{-1}$ and $10^{-2}$, for $N_c=1,3,5,7$. The dotted line
shows the asymptotic low-$|\omega|/T_K^{SC}$ FL behavior 
$D\rho_{K,L}(\omega)=1-a(|\omega|/T_K^{SC})^2$.
(C): For small asymmetry ($\rho\delta_K=10^{-6}$ and $10^{-7}$
for $N_c=1,3,5,7$), universal behavior arising on rescaling
$D\rho_{K,\alpha}(\omega)$ for \emph{both} $\alpha=L/R$ in terms of
$\omega/T_K^{2CK}$, as compared with the symmetric 2CK scaling curve
(circles). The dotted line shows the asymptotic ($T^{\star}\ll |\omega|
\ll T_K^{2CK}$) NFL behavior $D\rho_{K,\alpha}(\omega)=\tfrac{1}{2}[1-b(|\omega|/T_K^{2CK})^{1/2}]$.
(D): Same data rescaled in terms of $T/T^{\star}$, showing the
universal low-temperature crossover for $D\rho_{K,L}(\omega)$ (solid lines) and
$D\rho_{K,R}(\omega)\equiv 1-D\rho_{K,L}(\omega)$ (dashed). The 
$T^{\star}\ll |\omega| \ll T_K^{2CK}$ asymptotic
behavior $D\rho_{K,L}(\omega)= \tfrac{1}{2}[1 +
c(|\omega|/T^{\star})^{-1/2}]$ is shown as a dot-dashed line; while for
$|\omega|\ll T^{\star}$ FL behavior results,
$D\rho_{K,L}(\omega)= 1-d(|\omega|/T^{\star})^{2}$ (dotted line).
}
\end{figure}

\begin{figure}
\includegraphics[height=5cm]{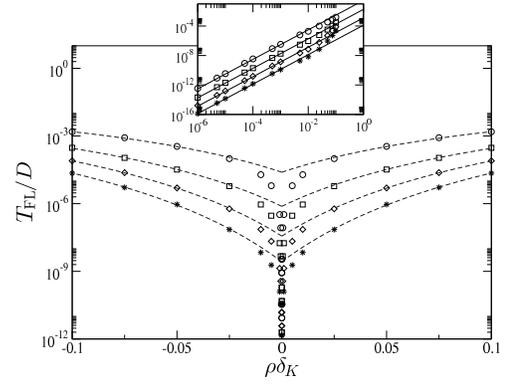}
\caption{\label{asym bigj tk}
Fermi liquid crossover temperature $T_{\text{FL}}/D$ \emph{vs} $\rho \delta_K$ for chains of length
$N_c=1,3,5,7$ (circles, squares, diamonds and stars)  and couplings $\rho
J=0.25$ and $\rho J_K=0.125$. For large channel
asymmetry, $T_{\text{FL}}\equiv T_K^{SC}$: good agreement between the data and
Eq.~\ref{SC asym pms} (dashed lines) is seen for each $N_c$ in the
regime $|\rho \delta_K|\gtrsim 0.025$. The inset shows the small asymmetry
behavior on a log-log plot. Here $T_{\text{FL}}\equiv T^{\star}$, and the
quadratic behavior of Eq.~\ref{asym scale van} is shown as the solid
lines, onto which data fall cleanly for $\rho|\delta_K|\lesssim0.01$.
}\end{figure}

In the case of smaller channel asymmetry,  $\rho \delta_{K}\ll \rho T_K^{1CK}$, RG flow to the stable
Fermi liquid SC:L FP occurs via the critical FP, which is of course
the 2CK FP. The chain spin-$\tfrac{1}{2}$ associated with the LM FP is then fully screened 
in a \emph{two-stage} process (Fig.~\ref{bigj asym ent fig}(A)). 
All such systems flow first to the 2CK FP on a  common scale $T_K^{2CK}$, 
given by Eq.~\ref{2CK pms}. The entropy thus drops to 
$S_{\text{imp}}=\tfrac{1}{2}\ln(2)$, symptomatic\cite{2ck:rev_cox_zaw} of overscreening; 
before being quenched completely below a scale $T\sim T^{\star}$
characterizing\cite{2ck:nozieres,2ck:rev_cox_zaw,2ck:BA_sacramento,2ck:BA_andrei_jerez,2ck:CFT_NRG} the
flow to the SC:L FP (with $T^{*}$ defined in practice by $S_{\text{imp}}(T^{*}) =\tfrac{1}{4}\ln2$).
A clear $S_{\text{imp}}=\tfrac{1}{2}\ln(2)$ plateau is thus seen for
lines (d)--(g) in Fig.~\ref{bigj asym ent fig}(A), with $T^{\star}$
diminishing rapidly as the transition is approached. The evolution of $T^{\star}$ 
on varying $\delta_K$ for different odd chains is itself studied in 
Fig.~\ref{asym bigj tk}, below; and the result in the small-$|\delta_K|$
regime is a characteristic power-law decay,
\begin{equation}
\label{asym scale van}
T^{\star}~\overset{\delta_K\rightarrow 0^{\pm}}\sim ~ {\cal{A}}|\delta_K|^{\nu}~~~~~~:\nu =2
\end{equation}
with exponent $\nu =2$, and common amplitudes $\cal{A}$ on approaching
the transition at $\delta_K=0$ from either side (as guaranteed by symmetry). 
Eq.~\ref{asym scale van} generalizes the known result\cite{2ck:rev_cox_zaw,2ck:BA_sacramento,2ck:BA_andrei_jerez} for the channel-anisotropic single-spin 2CK model, and is expected from the 
mapping of the full chain model (Eq.~\ref{Hfull}) onto the effective 
model, Eq.~\ref{Heff bigj}.

We now turn to the scaling behavior of the entropy for chains of
different length, as demonstrated by the three universal curves given
in panels (B)--(D) of Fig.~\ref{bigj asym ent fig}.  First,
in Fig.~\ref{bigj asym ent fig}(B) for $N_c=1,3,5,7$, we show strongly asymmetric 
systems ($\rho \delta_{K}\gg \rho T_K^{1CK}$) with $\rho\delta_K = 10^{-1}$ and $10^{-2}$.
The data clearly collapse to common scaling form when scaled in 
terms of $T/T_K^{SC}$, indicative of universal one-stage quenching from the LM to the SC:L FP.
Results for the single-channel spin-$\tfrac{1}{2}$ Kondo model are 
also shown (circles), confirming that the crossover is  
characterized by effective \emph{single-channel} Kondo screening. 

The situation is more subtle for weakly asymmetric systems $\rho \delta_{K}\ll \rho T_K^{1CK}$, where two-stage quenching occurs from the LM FP, through the 2CK FP,  to the fully quenched SC:L FP. As now shown, \emph{each} of these stages separately exhibit universal scaling, in terms of the two distinct low-energy scales $T_K^{2CK}$ and $T^{\star}$ respectively.
In Figs.~\ref{bigj asym ent fig}(C),(D) for $N_c=1,3,5,7$, systems close to the transition  
are shown. To determine the full universal curves,  it is of course essential to obtain good scale separation of $T_K^{2CK}$ and $T^{\star}$: here $T_K^{2CK}/T^{\star}>10^8$.

In Fig.~\ref{bigj asym ent fig}(C), results are rescaled in terms of
$T/T_K^{2CK}$. Collapse to the universal scaling curve for the
symmetric single-spin 2CK model (shown separately, circles) is seen clearly; the crossover from
the LM FP ($S_{\text{imp}}=\ln 2$) to the 2CK FP ($S_{\text{imp}}=\tfrac{1}{2}\ln 2$) being
as such determined by the 2CK scale $T_K^{2CK}$. By contrast, the universality of the crossover from the unstable 2CK FP to the stable low-$T$ SC:L FP with $S_{\text{imp}}=0$, is shown in Fig.~\ref{bigj asym ent fig}(D). Here, 
on rescaling in terms of $T/T^{\star}$ the data collapse to a universal form controlled by the low-energy scale $T^{*}$, itself vanishing (Eq.~\ref{asym scale van}) as the quantum critical point $\delta_{K}=0$ is approached. \\

The FP structure and energy scales naturally show up also in dynamical quantities, such as the scattering 
t-matrix, and  hence the spectra $D\rho_{K,\alpha}(\omega)$.
These are considered in Fig.~\ref{asym bigj spec}. In panel (A), again for $N_c=3$, we
focus on $D\rho_{K,L}(\omega)$ (solid lines) and 
$D\rho_{K,R}(\omega)$ (dashed) for systems with the same
parameters as Fig.~\ref{bigj asym ent fig}(A). Results for $\delta_K<0$
are not shown, the $\alpha=L$ and $R$ spectra simply being exchanged 
under the transformation $\delta_K \leftrightarrow -\delta_K$.

$N_{c}=3$ chains with $\rho\delta_K=10^{-1}$ and $10^{-2}$
[strong channel-asymmetry, lines (a) and (b)], show a characteristic resonance in
the left-channel spectrum, $D\rho_{K,L}(\omega)$, on the scale 
$|\omega|\sim T_K^{SC}$; with the Fermi level value in particular being $D\rho_{K,L}(0)=1$. This is the single-channel Kondo resonance; as is physically natural for these 
strongly channel-asymmetric cases because effective
single-channel Kondo screening is operative,\cite{2ck:CFT_NRG} with the behavior thus
expected to be that of the single-channel Kondo or Anderson models.\cite{hewson}
In the particle-hole symmetric limit of the latter, the Friedel sum rule\cite{lang}
guarantees satisfaction of the unitarity limit ($D\rho_{K,L}(0)=1$); and
in the scaling regime $|\omega|\ll J/N_c$ one expects the entire one-channel scaling spectrum to be recovered.
This is considered in panel (B), where  results for $N_c=1,3,5,7$ and $\rho\delta_K=10^{-1}$ and $10^{-2}$ 
are shown, rescaled in terms of $\omega/T_K^{SC}$: essentially
perfect agreement is seen with the universal scaling spectrum for the
single-channel Kondo model (shown separately as circles).
For $T_K^{SC} \ll |\omega| \ll J/N_c$ the characteristic
$D\rho_{K,L}(\omega)\sim A/[\ln^2(|\omega|/T_K^{SC})+B]$
behavior typical of `high' energy spin-flip scattering\cite{NLDscaling,LMA:SU(2N)} arises; 
while for $|\omega|\ll T_K^{SC}$ canonical Fermi
liquid behavior,\cite{hewson}  $D\rho_{K,L}(\omega)= 1-a(|\omega|/T_K^{SC})^2$, 
occurs as expected.

Spectra for the right lead/channel, $D\rho_{K,R}(\omega)$ [dashed
lines (a) and (b) of panel (A)], are similarly described by the 
leading $\sim 1/\ln^2(|\omega|/T_K^{SC})$ logarithms at high
energies. However, the upward renormalization of the effective Kondo coupling 
to the right lead --- and hence RG flow towards the SC:R FP --- is cut
off at $|\omega|\sim T_K^{SC}$, below which frequency the impurity chain
local moment becomes fully screened by strong coupling to the
\emph{left} lead.
Thus $D\rho_{K,R}(\omega)=0$ for $|\omega|\ll
T_K^{SC}$, as observed directly from the NRG results in panel (A).

We now turn to lines (e)--(g) of Fig.~\ref{asym bigj spec}(A), 
for $N_c=3$ systems with much smaller channel-asymmetry, $\rho\delta_{K}\ll \rho T_{K}^{1CK}$.
For both $\alpha=L$ and $R$, a clear half-unitary
plateau of $D\rho_{K,\alpha}(\omega)\simeq\tfrac{1}{2}$ arises for
$T^{\star}\ll |\omega| \ll T_K^{2CK}$, indicative of RG flow near
the 2CK FP. For $|\omega| \sim T^{\star}$, however, flow to the Fermi
liquid SC:L FP occurs, such that $D\rho_{K,L}(\omega=0)=1$
and $D\rho_{K,R}(\omega=0)=0$ are again satisfied. 
As was seen from the entropy (Fig.~\ref{bigj asym ent fig}), 
there are two universal scales in this
regime associated with the crossover from the LM FP to the 2CK FP [see
panel (C)] and from the 2CK FP to the SC:L FP [panel (D)].

In panel (C) of Fig.~\ref{asym bigj spec}, results are shown for systems of chain length
$N_c=1,3,5,7$ and small channel asymmetry, $\rho\delta_K=10^{-6}$ and
$10^{-7}$. Each is rescaled in terms of $\omega/T_K^{2CK}$, and
collapse to the universal \emph{symmetric} 2CK curve (circles) is seen in all
cases --- for \emph{both} $\alpha=L$ and $R$ spectra. In particular, 
for $|\omega|\ll T_K^{2CK}$ the characteristic NFL behavior is obtained, 
$D\rho_{K,\alpha}(\omega)= \tfrac{1}{2}[1-b(|\omega|/T_K^{2CK})^{1/2}]$ (dotted line).

Finally, panel (D) shows the same data, but rescaled now in terms
of $|\omega|/T^{\star}$. Two universal spectra emerge: one for 
$D\rho_{K,L}(\omega)$ and one for
$D\rho_{K,R}(\omega)$. The two scaling spectra are however found to be related simply by 
$D\rho_{K,R}(\omega)=1-D\rho_{K,L}(\omega)$, so we need
consider only $D\rho_{K,L}(\omega)$ (solid line). 
For $\omega\gg T^{\star}$ the relevant $L/R$
symmetry-breaking operator dominates,\cite{2ck:CFT_NRG} 
driving RG flow away from the 2CK FP. Since the scaling dimension 
of this operator\cite{2ck:CFT_NRG} is $\tfrac{1}{2}$,
one expects $D\rho_{K,L}(\omega)= \tfrac{1}{2}[1 +
c(|\omega|/T^{\star})^{-1/2}]$ (as indeed found, dot-dashed line). By contrast, for 
$|\omega|\ll T^{\star}$, irrelevant operators\cite{KWW,2ck:CFT_NRG} 
affect the RG flow in the vicinity of the stable FL FP, so one expects the leading low-$|\omega|/T^{\star}$
asymptotics to be
$D\rho_{K,L}(\omega)= 1-d(|\omega|/T^{\star})^{2}$
(dotted line). Good agreement with the numerics is seen in both regimes and for both
$\alpha=L/R$ spectra in Fig.~\ref{asym bigj spec}(D).

As the critical point is approached [$|\delta_K|\rightarrow 0$; 
lines (a)$\rightarrow$(h) in Fig.~\ref{asym bigj spec}], the spectra fold progressively onto the
2CK spectrum itself [line (h)] down to lower and lower
frequencies. The scale $T^{\star}$ describing flow away from the 2CK
FP vanishes according to Eq.~\ref{asym scale van}, as evident from the
dynamics shown in Fig.~\ref{asym bigj spec} (or the thermodynamics in  
Fig.~\ref{bigj asym ent fig}). In Fig.~\ref{asym bigj tk} the evolution of
the low-energy scale as a function of channel asymmetry, $\rho\delta_K$, is
examined. Since $T_K^{SC}$ is the lowest energy scale of the problem
in the large-$|\delta_K|$ regime, while $T^{\star}$ is the lowest scale
for small $|\delta_K|$, we consider the generic crossover scale $T_{\text{FL}}$ in
Fig.~\ref{asym bigj tk}, defined in practice from the entropy via
$S_{\text{imp}}(T_{\text{FL}})=\tfrac{1}{4}\ln(2)$; which as such characterizes
the flow to the ultimate stable FL FP, and hence 
complete screening of the impurity spin. $T_{\text{FL}}/D$ is
shown \emph{vs} $\rho \delta_K$ for chains of length $N_c=1,3,5,7$. 
The dashed lines in the main panel show
comparison to the perturbative result for $T_K^{SC}$ given in 
Eq.~\ref{SC asym pms} (with a prefactor $\mathcal{O}(1)$ 
for each $N_c$ adjusted to fit the numerics). The inset shows the same 
data on a log-log scale, demonstrating the quadratic decay of
$T^{\star}$, given by Eq.~\ref{asym scale van} in the
small-$|\delta_K|$ regime (solid lines).


\subsection{Weak inter-impurity coupling}\label{smallJ}
The perturbative derivation of the effective 2CK model in Sec.~\ref{largeJ eff} is valid for sufficiently large inter-impurity exchange couplings, in which regard we note that any bare energy scale
larger than the the exponentially-small universal scales $T_K^{2CK}$ or
$T_{\text{FL}}$ may be considered `large'.

However in the $J=0$ limit, the physical behavior is clearly very different. Here,  each terminal
spin-$\tfrac{1}{2}$ impurity undergoes the standard
\emph{single-channel} Kondo effect\cite{hewson} with its attached $\alpha=L/R$ lead below a temperature $T\sim
T_{K,\alpha}^{1CK}$, while the
remaining impurities remain free down to $T=0$. This scale is
associated with the flow to the strong coupling (SC) FP,\cite{hewson} and is given
from perturbative scaling\cite{hewson} by
\begin{equation}
\label{1CK pms}
T_{K,\alpha}^{1CK}\sim D\sqrt{\rho J_{K\alpha}} \exp(-1/\rho J_{K\alpha} ).
\end{equation}
The question addressed in this section is: what is the physics for
odd chains with small but finite AF inter-impurity 
coupling, $ J \lesssim T_{K,\alpha}^{1CK}$?

A physically intuitive picture for the simplest $N_c=3$ case is
depicted schematically in Fig.~\ref{smallj schematic} (and discussed
further in Sec.~\ref{smallJ eff} below). Impurity `1'
forms a single-channel Kondo singlet with the left lead,
and impurity `3' likewise with the right lead. A Fermi liquid 
description then applies, and the remaining states of each lead act
as an effective bath of non-interacting electrons that 
participate in the screening of impurity `2'. A residual  
AF exchange coupling, mediated via the Kondo singlets, 
once again yields an effective 2CK model. 

The above scenario is supported by the behavior of
single-channel systems involving side-coupled quantum
dots.\cite{side:boese,side:cornaglia,side:zb,fanokondoDD} In the
simplest example of a dot dimer, two-stage Kondo
screening is operative in the small inter-dot coupling
regime\cite{side:cornaglia}: the dot connected to the lead undergoes
a spin-$\tfrac{1}{2}$ Kondo effect on the scale $T_{K,1}$,
while residual AF coupling between the remaining dot and the lead
gives rise to a second Kondo effect for $T\sim T_{K,2}$ ($\ll
T_{K,1}$), leading thereby to complete screening of both dots on the
lowest energy scales.\cite{side:cornaglia}  Similar mechanisms
have been advanced to describe the low-energy behavior of an
asymmetric two-channel two-impurity Kondo
model\cite{2ck:chains_zarand} and a triple quantum dot ring
structure.\cite{zitkoTQD2ch}  

In the present context of odd impurity chains coupled to two leads,
the most interesting behavior is expected in the $L/R$-symmetric
case. Here the effective coupling to left and right leads is also
symmetric, and hence the 2CK FP must describe the low-energy behavior of
the system.


\begin{figure}
\includegraphics[height=2.5cm]{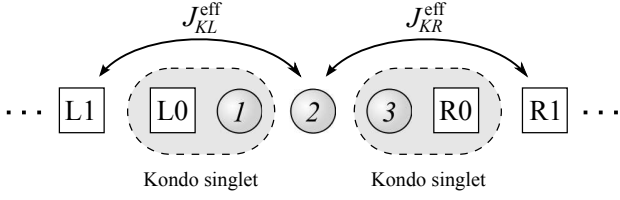}
\caption{\label{smallj schematic}
Schematic illustration for the $N_c=3$ chain with small inter-impurity
coupling $J\lesssim T_{K,\alpha}^{1CK}$, as discussed in text. 
}\end{figure}

\subsubsection{Effective 2CK model for $N_c=3$}\label{smallJ eff}
Before considering  NRG calculations, we first derive the effective
2CK model for the simplest $N_c=3$ case; using perturbative techniques and 
scaling arguments, and exploiting the Wilson chain representation\cite{nrgreview,KWW,nrg_rev}
(see Fig.~\ref{smallj schematic}) as natural within an RG framework.

The Wilson chain for lead $\alpha=L/R$ is defined\cite{nrgreview,KWW,nrg_rev} by dividing the band
up into logarithmic intervals,
$\{\pm D \Lambda^{-n}\}$ ($n=0,1,2,..$), and then discretizing it by retaining only the
symmetric combination of states within each interval. This
Hamiltonian is then tridiagonalized to obtain a linear chain form,
with the impurity system coupled at one end.\cite{nrgreview,KWW,nrg_rev} 
The $N_c=3$ Hamiltonian may thus be written in a dimensionless form $H_N=H_0+H_1$, 
\begin{equation}
\label{smallj nrg ham}
\begin{split}
H_0 &= \mathcal{J}_{KL} \hat{\textbf{S}}_1\cdot \hat{\textbf{s}}_{L0}+\mathcal{J}_{KR} \hat{\textbf{S}}_3\cdot \hat{\textbf{s}}_{R0}\\
H_1 &= \mathcal{J} (\hat{\textbf{S}}_1\cdot
\hat{\textbf{S}}_2+\hat{\textbf{S}}_2\cdot \hat{\textbf{S}}_3) \\
&+  \sum_{\alpha,\sigma} \sum_{n=0}^{N-1}t_n\left ( f_{\alpha n \sigma}^{\dagger} f_{\alpha (n+1) \sigma}^{\phantom{\dagger}} +  f_{\alpha (n+1) \sigma}^{\dagger} f_{\alpha n \sigma}^{\phantom{\dagger}} \right )
\end{split}
\end{equation}
where $\hat{\textbf{s}}_{\alpha 0}$ is given in Eq.~\ref{0 orb} and the Wilson chain operators $f_{\alpha n \sigma}^{\dagger}$ are obtained recursively using the Lanczos algorithm.\cite{nrgreview} 
The rescaled dimensionless couplings are given by 
$\mathcal{J}_{K\alpha}=2\rho J_{K\alpha}/A_N$ and $\mathcal{J}=2\rho
J/A_N$ (where $A_N=\tfrac{1}{2}(1+\Lambda^{-1})\Lambda^{-(N-1)/2}$).
For a flat-band lead density of states, the tunnel-coupling between Wilson chain orbitals 
takes the form\cite{nrgreview} $t_n=\Lambda^{(N-1)/2}\Lambda^{-n/2}\xi_n$, with the
$\xi_n\equiv \xi_n(\Lambda)\sim \mathcal{O}(1)$. The full Hamiltonian is then recovered in the $N\rightarrow \infty$
limit\cite{nrgreview,KWW,nrg_rev} via
$H={\text{lim}}_{N\rightarrow\infty} \left \lbrace DA_N H_N \right \rbrace$.

We consider first the limit of strong impurity-lead coupling,
$\mathcal{J}_{K\alpha}\gg \max(\mathcal{J}, t_n)$; so that $H_0$ in
Eq.~\ref{smallj nrg ham} favours formation of a pair of singlet states between the terminal impurities and the `0'-orbital of their attached lead. The ground state of $H_0$ thus comprises a 1-`L0' singlet (we denote it by $|s;L\rangle$) and a 3-`R1' singlet (denoted $|s;R\rangle$), as shown schematically in Fig.~\ref{smallj schematic}.

$H_1$ now acts perturbatively, and we project onto the lowest
state of $H_0$ using the unity operator for the reduced Hilbert space,
$\hat{1}_{s}=|s;L\rangle|s;R\rangle\langle s;R|\langle s;L|$.
An effective Hamiltonian may be obtained using the Brillouin-Wigner
perturbation expansion,\cite{ziman} $H_N=E_0+H^{\text{eff}}_{\text{I}}
+ H^{\text{eff}}_{\text{II}} + H^{\text{eff}}_{\text{III}}+..$. 
Here, $E_0=\hat{1}_s H_0 
\hat{1}_s = -\tfrac{3}{4}(\mathcal{J}_{KL}+\mathcal{J}_{KR})$ is merely
a constant shift in energy, while $H^{\text{eff}}_{\text{I}}=\hat{1}_{s} H_1 \hat{1}_{s}\equiv\tilde{H}_L$ follows as
\begin{equation}
\label{0orb removed}
\tilde{H}_L =  \sum_{\sigma, \alpha} \sum_{n=1}^{N-1}t_n\left ( f_{\alpha n \sigma}^{\dagger} f_{\alpha (n+1) \sigma}^{\phantom{\dagger}} +  f_{\alpha (n+1) \sigma}^{\dagger} f_{\alpha n \sigma}^{\phantom{\dagger}} \right )
\end{equation}
and corresponds to a pair of free Wilson
chains, with the `0'-orbital of each removed. The second-order term,
$H^{\text{eff}}_{\text{II}}$ contributes only potential scattering\cite{hewson},
here omitted for clarity. An effective coupling between impurity `2' and the `L1' and `R1' orbitals is generated only to \emph{third}-order in $H_1$, given\cite{ziman} by
$H^{\text{eff}}_{\text{III}}=\hat{1}_{s} H_1 (E_0-H_0)^{-1}\hat{P}H_1(E_0-H_0)^{-1} \hat{P}H_1 \hat{1}_{s}$
with $\hat{P}=\hat{1}-\hat{1}_{s}$ a projector. Combining this with Eq.~\ref{smallj nrg ham},
a rather lengthy calculation yields
\begin{equation}
\label{Heff smallj odd}
H^{\text{eff}}_{\text{III}}= \rho J_{KL}^{\text{eff}} \hat{\textbf{S}}_2\cdot \hat{\textbf{s}}_{L1}+\rho J_{KR}^{\text{eff}} \hat{\textbf{S}}_2\cdot \hat{\textbf{s}}_{R1}
\end{equation}
(omitting RG irrelevant terms), with the effective coupling of impurity `2' to the 
$\alpha=L$/$R$ lead given by 
\begin{equation}
\label{eff coup smallj}
\rho J_{K\alpha}^{\text{eff}}= \left ( \frac{20 t_0^2\mathcal{J}}{9 \mathcal{J}_{K\alpha}^2}\right )>0.
\end{equation}
The effective model 
$H_N^{\text{eff}}=\tilde{H}_L +H^{\text{eff}}_{\text{III}}$, describes thereby the residual 
AF coupling between impurity `2' and the `L1' and `R1' 
orbitals of a pair of leads (with impurities `1' and `3' and lead 
orbitals `L0' and `R0' removed); see Fig.~\ref{smallj schematic}. This is a model of 2CK form.

The above analysis presupposes the existence of the local
singlet states $|s;L\rangle$ and $|s;R\rangle$. However, we note that
for $J\lesssim T_{K,\alpha}^{1CK}$ (as given by Eq.~\ref{1CK pms}), RG
flow is expected near a Fermi liquid-type FP, comprising 
single-channel strong coupling states in each lead, with a free,
disconnected local moment on impurity `2' (and which `SC x SC x LM' FP is of course stable only at
the point $J=0$). Renormalization of the impurity-lead coupling 
$J_{K\alpha} \rightarrow \tilde{J}_{K\alpha}$ on successive 
reduction of the temperature/energy scale naturally results in 
incipient formation of Kondo singlets between each 
terminal impurity and its attached lead below 
$T\sim T_{K,\alpha}^{1CK}$ for the $\alpha=L/R$ channel,
respectively.  An effective 2CK model should then result via the
mechanism described above, where the local singlet states 
$|s;L\rangle$ and $|s;R\rangle$ are now \emph{Kondo} singlets.

The central question then is: how does 
Eq.~\ref{eff coup smallj} flow under renormalization? 
Specifically, what is the effective coupling 
$\rho \tilde{J}_{K\alpha}^{\text{eff}}$ for $T\sim T_{K,\alpha}^{1CK}$?

To answer this, recall first that the effective 
temperature\cite{nrgreview,KWW,nrg_rev} within
the RG framework is related to the iteration number/Wilson
chain length, $N$, via $T\sim \Lambda^{-N/2}$. The operators for the
Wilson chain orbitals also scale with $N$. In particular,
operators for the `0'-orbital of the Wilson chain 
scale as\cite{nrgreview,KWW} $f_{\alpha 0 \sigma}\sim
\Lambda^{-N/4}$. Thus  $\mathcal{J}_{K\alpha}\rightarrow
\tilde{\mathcal{J}}_{K\alpha}\sim \Lambda^{-N/2}$,  since the
impurity-lead exchange coupling is associated with a pair of
`0'-orbital fermionic operators. The key result is thus that the
renormalized impurity-lead coupling $\tilde{\mathcal{J}}_{K\alpha}\sim
T_{K,\alpha}^{1CK}$ for $T\sim T_{K,\alpha}^{1CK}$ --- in accord with the physical
expectation that disruption of the $\alpha=L/R$ Kondo singlet costs an
energy $\mathcal{O}(T_{K,\alpha}^{1CK})$. By contrast, the coupling
between the impurities, $ \mathcal{J}$, is not associated with any 
chain operators, and hence does \emph{not} get renormalized with $N$.  Further, as pointed out
in Ref.~\onlinecite{chen+jay}, once the `0'-orbital of a Wilson chain has been frozen out 
(e.g.\ by formation of a Kondo singlet), the `1'-orbital operators then scale as
$f_{\alpha 1 \sigma}\sim \Lambda^{-N/4}$. Thus the renormalized
tunnel-coupling $\tilde{t}_0\sim \Lambda^{-N/4}$, so that for $T\sim
T_{K,\alpha}^{1CK}$, $\tilde{t}_0^2/\tilde{\mathcal{J}}_{K\alpha}$ remains
$\mathcal{O}(1)$.  From Eq.~\ref{eff coup smallj}, the renormalized effective Kondo coupling
at $T\sim T_{K,\alpha}^{1CK}$ can then be estimated to have the
functional dependence $\rho\tilde{J}_{K\alpha}^{\text{eff}} \sim J/T_{K,\alpha}^{1CK}$.

For simplicity we focus now on the mirror symmetric case, where $J_{K\alpha}\equiv J_K$ and 
$T_{K,\alpha}^{1CK}\equiv T_K^{1CK}$, whence one has the effective low-energy Hamiltonian for $N_c=3$
\begin{equation}
\label{Heff smallj}
H^{\text{eff}}_{N_c}= \tilde{H}_L+\rho \tilde{J}_{K,N_c}^{\text{eff}}
\hat{\textbf{S}}\cdot (\hat{\textbf{s}}_{L1} +\hat{\textbf{s}}_{R1}),
\end{equation}
with $\hat{\textbf{S}} \equiv \hat{\textbf{S}}_2$ and 
the effective coupling
\begin{equation}
\label{renorm coup smallj}
\rho \tilde{J}_{K,N_c}^{\text{eff}} = \left (
  \frac{J}{T_{K}^{1CK}}\right )x(N_c)
\end{equation}
valid for $T\lesssim T_K^{1CK}$. Determination of the
constant $x(N_c)$ is obviously beyond the scope of this analysis,
although it can be deduced directly from NRG calculations as
demonstrated in the next section. 

2CK physics is thus expected for $T\sim T_K^{2CK}$ ($\ll T_K^{1CK}$),
as given from perturbative scaling\cite{2ck:nozieres} by
\begin{equation}
\label{2CK pms smallJ}
T_K^{2CK}\sim T_K^{1CK}\rho \tilde{J}_{K,N_c}^{\text{eff}} \exp(-1/\rho \tilde{J}_{K,N_c}^{\text{eff}} ),
\end{equation}
where the physical origin of the prefactor $T_K^{1CK}$ is simply that
the effective bandwidth of the problem is already reduced to $\sim
T_K^{1CK}$ at the temperature $T\sim T_K^{1CK}$, below which the effective model,
Eq.~\ref{Heff smallj}, is valid.


\begin{figure}
\includegraphics[height=5cm]{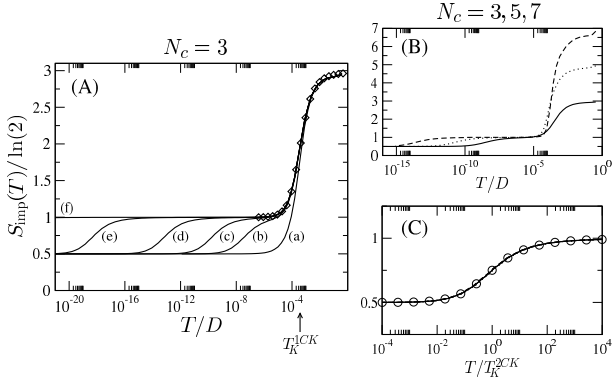}
\caption{\label{smallj ent fig}
(A): 
$S_{\text{imp}}(T)/\ln(2)$ \emph{vs} $T/D$ as the inter-impurity
coupling is reduced, for the symmetric $N_c=3$ case ($\rho \delta_K=0$). Shown for fixed 
$\rho J_K=0.15$, varying $\rho J=\lambda (T_K^{1CK}/D)$,
where $\lambda= 10^{-n/10}$ and $n=0, 5, 7, 9, 11, 13$ [lines
(a)--(f)], with $T_K^{1CK}$ determined from a $J=0$
calculation. Behavior for $T\gg T_K^{2CK}$ is described by 
$S^{J=0}_{\text{imp}}(T)=\ln(2)+2S^{1CK}_{\text{imp}}(T)$ (diamonds), with
$S^{1CK}_{\text{imp}}(T)$ for a single-channel Kondo
model with the same $J_K$.
(B): Comparison of behavior for $N_c=3,5,7$ (solid, dotted and dashed
lines) with $J = \tfrac{1}{2}T_K^{1CK}$.
(C): Results for $N_c=3,5,7$ (with $n=9$ and $11$):
scaling collapse to the universal 2CK curve (circles) is seen in all cases. 
}\end{figure}

\subsubsection{NRG results for odd chains}\label{smallJ nrg}
The physical picture for the $N_c=3$ system is thus clear, and
we now turn to NRG results for odd chains of
length $N_c=3,5,7$ in the regime of weak coupling between the
impurities. For accurate numerics, we found it necessary to
retain $N_s=4000$, $6000$ and $12000$ states per iteration for
$N_c=3$, $5$ and $7$, since higher-energy chain states remain
important down to $T\sim J\lesssim T_K^{1CK}$.

Fig.~\ref{smallj ent fig}(A) shows $S_{\text{imp}}(T)$ \emph{vs} $T/D$
for the $N_c=3$ case discussed explicitly above; for a common $\rho J_K$ and with 
$\rho J=\lambda (T_K^{1CK}/D)$, where $\lambda= 10^{-n/10}$ and $n=0, 5, 7, 9, 11, 13$ [lines
(a)--(f)]. $T_K^{1CK}$ was itself determined from a $J=0$ calculation
which, modulo a free spin on impurity 2, is equivalent to two
separate single-channel Kondo models with the same $J_K$. At high $T$,
the trivial $S_{\text{imp}}=3\ln(2)$ behavior expected for three free
spins-$\tfrac{1}{2}$ arises in all cases. Line (a) crosses
directly to $S_{\text{imp}}=\tfrac{1}{2}\ln(2)$ on the scale
$J=T_K^{1CK}\approx  T_K^{2CK}$, characterizing flow to the 2CK FP. By contrast, 
lines (b)--(f) flow first to the SC x SC x LM FP ($S_{\text{imp}}=\ln(2)$). We also show for comparison  
$S^{J=0}_{\text{imp}}(T)=\ln(2)+2S^{1CK}_{\text{imp}}(T)$ (diamonds), where
$S^{1CK}_{\text{imp}}(T)$ is the entropy of a single-channel Kondo
model\cite{hewson} with the same Kondo coupling; $S^{J=0}_{\text{imp}}(T)$ thus describes
the \emph{entire} temperature-dependence of the entropy 
for $J=0$. Lines (c)--(f) follow this curve perfectly for $T\gg
T_K^{2CK}$, as expected from the single-channel Kondo
screening of impurities `1' and `3'. An intermediate $S_{\text{imp}}=\ln(2)$ 
plateau is thus observed, the single-channel $T_K^{1CK}$ remaining constant while 
the two-channel scale $T_K^{2CK}$ diminishes rapidly as $J$ is decreased.
RG flow thus persists in the vicinity of the SC x SC x LM FP over an extended
$T$-range, but below $T\sim T_K^{2CK}$ all systems are
described by the 2CK FP, with residual entropy
$S_{\text{imp}}=\tfrac{1}{2}\ln(2)$.  

\begin{figure}
\includegraphics[height=5cm]{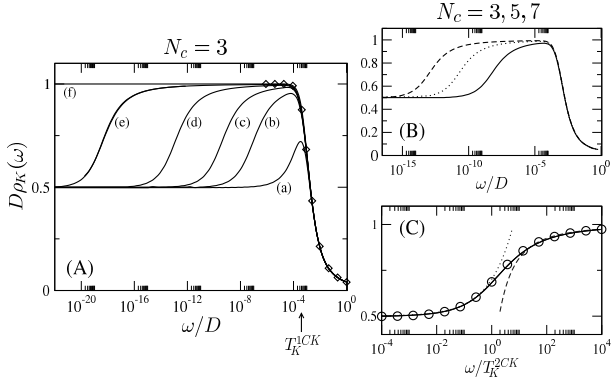}
\caption{\label{smallj dyn fig}
(A): Spectra $D\rho_K(\omega)$ \emph{vs} $\omega/D$ as the inter-impurity
coupling is reduced, for the symmetric $N_c=3$ case (with same
parameters as Fig.~\ref{smallj ent fig}). Behavior at 
$|\omega| \gg T_K^{2CK}$ is described by $D\rho_K^{1CK}(\omega)$ (diamond
points) for a single-channel Kondo model.
(B): Comparison of $N_c=3,5,7$ (solid, dotted, dashed
lines) with $J = \tfrac{1}{2}T_K^{1CK}$.
(C): Spectra \emph{vs} $\omega/T_K^{2CK}$ for $N_c=3,5,7$ with $n=9$ and $11$:
universal scaling collapse is seen. For $T_K^{2CK}\ll
|\omega|\ll T_K^{1CK}$ the asymptotic behavior is 
$D\rho_K(\omega)=1-A/[\ln^2(|\omega|/T_K^{2CK})+B]$ (grey dashed line), while for
$|\omega|\ll T_K^{2CK}$ NFL scaling is observed: 
$D\rho_{K}(\omega) = \tfrac{1}{2}[1+b(|\omega|/T_K^{2CK})^{1/2}]$
(grey dotted line). The entire curve is described by 
$D\rho_K(\omega)=1-D\rho_K^{2CK}(\omega)$ (circles), with $\rho_K^{2CK}(\omega)$
the scaling spectrum of the standard 2CK model.
}\end{figure}

\begin{figure}
\includegraphics[height=5cm]{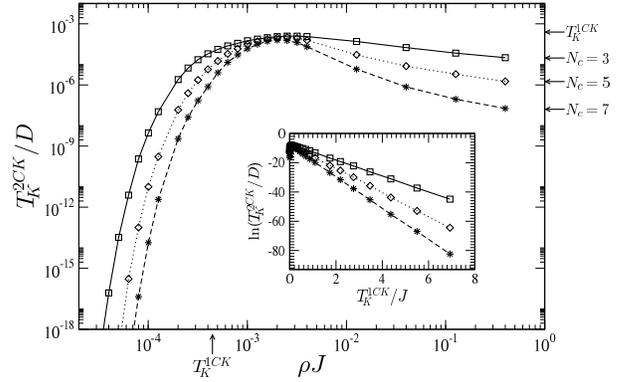}
\caption{\label{smallj tk fig}
Evolution of the two-channel Kondo scale: $T_K^{2CK}/D$ \emph{vs} $\rho J$
for symmetric chains of length $N_c=3,5,7$ (squares, diamonds and stars), 
and Kondo coupling $\rho J_K = 0.15$ (with $\delta_K=0$). 
For large inter-impurity coupling,  $T_K^{2CK}$ is
given by Eqs.~\ref{eff coup bigj},\ref{2CK pms} (indicated by
arrows). For each chain, the maximum value $T_K^{2CK}\approx
T_K^{1CK}$ is obtained for $J \approx 10 T_K^{1CK}$. For $J <
T_K^{1CK}$, the Kondo scale diminishes rapidly; as seen also in
the inset, where $\ln(T_K^{2CK})$ \emph{vs} $T_K^{1CK}/J$ is shown,
confirming Eqs.~\ref{renorm coup smallj} and \ref{2CK pms smallJ}. 
The slopes yield $x(N_c=3)\approx 1/6$, $x(N_c=5)\approx 1/9$ and $x(N_c=7)\approx 1/12$.
}\end{figure}

We now comment on the generic behavior expected for
impurity chains with $N_c>3$, which is a physically natural extension of 
the $N_{c} =3$ case above. Following the `removal' of the terminal 
impurities through formation of single-channel Kondo singlets
for $T\sim T_K^{1CK}$, the remaining odd $(N_c-2)$ impurities
form a residual spin-$\tfrac{1}{2}$ on the scale $T\sim J$
($\lesssim T_K^{1CK}$). This doublet state now feels an effective coupling to
the two leads via the mechanism described in Sec.~\ref{smallJ eff}, but with a further
renormalization of the effective Kondo exchange, as expected from the discussion in
Sec.~\ref{largeJ} in the regime of large inter-impurity coupling. 
Extension of the above analysis for $N_{c}=3$, which we do not give here, then leads us
to expect (as tested below, Fig.~\ref{smallj tk fig}) that the form Eq.~\ref{renorm coup smallj} should hold for odd $N_{c} >3$, with ratios $x(N_{c}+2)/x(N_{c}=3)$ which are the same as those inferred from Table~\ref{table:coup} but with  two sites excluded from the spin-chain (reflecting quenching of the terminal spins `1' and `$N_{c}$' to form Kondo singlets); i.e.\ from Table~\ref{table:coup} 
that $x(N_c=5)\approx \tfrac{2}{3} x(N_c=3)$ and $x(N_c=7)\approx 0.51 x(N_c=3)$.

2CK physics is thus is expected for all odd chains in the small inter-impurity coupling regime below
$T_K^{2CK}$, as given by Eqs.~\ref{2CK pms smallJ},\ref{renorm coup smallj}.
This is confirmed in Fig.~\ref{smallj ent fig}(B), where we consider 
chains of length $N_c=3,5,7$, taking $J=\tfrac{1}{2} T_K^{1CK}$ as an illustrative example.
Again, at the highest temperatures $T\gg J$, one obtains $S_{\text{imp}}=N_c\ln(2)$. For sufficiently large 
separation between $T_K^{1CK}$ and $J$, one expects the entropy to drop first to 
$S_{\text{imp}}=(N_c-2)\ln(2)$ on the scale $T_K^{1CK}$ (due to single-channel Kondo quenching\cite{hewson} of the terminal impurities), followed by a further drop for $T\sim J$ to the LM value
$S_{\text{imp}}=\ln(2)$; although in Fig.~\ref{smallj ent fig}(B) $T_K^{1CK}$ and $J$
are comparable, so no distinct $(N_c-2)\ln(2)$ plateau arises. In all cases, however,
$S_{\text{imp}}=\tfrac{1}{2}\ln(2)$ is seen below $T_K^{2CK}$,
characteristic of flow to the stable 2CK FP;\cite{2ck:rev_cox_zaw} with scales
$T_K^{2CK}$ that evidently diminish with increasing $N_c$, as expected qualitatively from
the above discussion.
Fig.~\ref{smallj ent fig}(C) shows the scaling curve
obtained when results are plotted \emph{vs} $T/T_K^{2CK}$ (with $n=9, 11$ chosen
to ensure good scale separation between $T_K^{2CK}$ and $T_K^{1CK}$).
The universal curve is precisely that of the standard 2CK model (circles). \\

Dynamics are now considered briefly, Fig.~\ref{smallj dyn fig} (A) showing
the $T=0$ $D\rho_K(\omega)$ for $N_{c} =3$ chains with the same parameters as in Fig.~\ref{smallj ent fig}(A). 
All systems show RG flow in the vicinity of the Fermi liquid-type SC x SC x LM FP (reflecting single-channel
Kondo screening of the terminal impurities), and hence an incipient
single-channel Kondo resonance\cite{hewson} in each channel. For comparison, we show 
$D\rho^{1CK}_K(\omega)$ for a standard single-channel Kondo model with
the same Kondo coupling (diamonds), which recovers perfectly the spectral
behavior for $|\omega|\gg T_K^{2CK}$. In particular, 
for lines (d)--(f) in the range $T_K^{2CK}\ll |\omega| \ll
T_K^{1CK}$, characteristic FL behavior\cite{hewson} $D\rho_K(\omega)=1-a(|\omega|/T_K^{1CK})^2$
arises,  and thus the unitarity limit $D\rho_K(\omega)=1$ is reached in this intermediate energy
window. For any finite $J$, $T_K^{2CK}$ is however always finite, so ultimately RG flow to the stable 2CK FP yields 
$D\rho_K(\omega)=\tfrac{1}{2}$ for $|\omega|\ll T_K^{2CK}$.

Fig.~\ref{smallj dyn fig}(B) shows spectra for 
$N_c=3,5,7$ chains with common 
$J = \tfrac{1}{2}T_K^{1CK}$, as in Fig.~\ref{smallj ent fig}(B).
These display the same qualitative behavior as for $N_c=3$, with a single-channel
Kondo resonance appearing at $|\omega|\sim T_K^{1CK}$, before crossing
over to the 2CK FP for $|\omega|\sim T_K^{2CK}$. Since $T_K^{2CK}$
diminishes with increasing chain length, while $T_K^{1CK}$
remains fixed, apparent FL behavior consequently persists down to
lower energies for the longer chains.

For $N_c=3,5,7$ chains with the same couplings as in Fig.~\ref{smallj ent fig}(C), Fig.~\ref{smallj dyn fig}(C) shows the spectrum arising when results are shown \emph{vs} $\omega/T_K^{2CK}$: collapse to a single
universal scaling curve is seen clearly. 
One might naively expect to obtain the  2CK scaling
spectrum here, since in this regime an effective 2CK model 
(Eq.~\ref{Heff smallj}) describes the system. 
However this model is only valid after the
single-channel Kondo effect has already taken place in each lead, conferring a
phase shift of $\pi/2$ to the conduction electrons.\cite{hewson} In
consequence, the scaling spectrum in the small $J$ limit is
$D\rho_K(\omega)=1-D\rho_K^{2CK}(\omega)$, with
$D\rho_K^{2CK}(\omega)$ the scaling spectrum of the
regular single-spin 2CK model. Comparison with the latter
(circles) confirms this directly.

Finally, in Fig.~\ref{smallj tk fig} we analyze the variation of the two-channel Kondo
scale as a function of the inter-impurity coupling strength, $\rho J$, and the
chain length, $N_{c}$. The $T_K^{2CK}/D$ in the large-$J$ limit
(indicated by arrows) are in accord with 
Eqs.~\ref{eff coup bigj},\ref{2CK pms}. However, as $\rho J$
is decreased, $T_K^{2CK}$ first increases (reaching its maximum of
$T_K^{2CK} \approx T_K^{1CK}$ for $J \approx 10 T_K^{1CK}$ in
each case), then diminishes very rapidly for $J\ll T_K^{1CK}$.
The behavior for small inter-impurity coupling is seen most clearly
in the inset, where $\ln(T_K^{2CK}/D)$ is shown
\emph{vs} $T_K^{1CK}/J$. The linear behavior confirms Eqs.~\ref{renorm coup
  smallj},\ref{2CK pms smallJ}, with the slopes yielding 
$x(N_c=3)\approx 1/6$, $x(N_c=5)\approx 1/9$ and
$x(N_c=7)\approx 1/12$, as consistent with the expectation
discussed above.


\begin{figure}
\includegraphics[height=3cm]{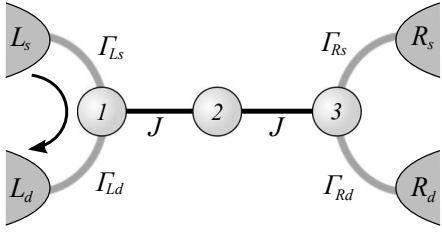}
\caption{\label{cond schematic}
Schematic of the $N_c=3$ trimer with impurities `1' and `3' modelled
as Anderson-like correlated levels, each tunnel-coupled to both source
($s$) and drain ($d$) leads, with  hybridizations
$\Gamma_{\alpha\gamma}$ (where $\alpha=L/R$ and $\gamma = s/d$).
The impurities are exchange-coupled to eliminate $L \leftrightarrow R$ charge
transfer processes that destroy 2CK physics, but conductance 
through impurity 1(3) in channel $\alpha=L(R)$ can be measured.
}\end{figure}

\section{Gate voltage effects and conductance}\label{GateV}
Having analyzed in detail the Heisenberg chain model,
Eq.~\ref{Hfull}, we now consider a variant in which the terminal
impurities are treated as correlated \emph{levels}
(dots), tunnel-coupled to their respective leads. Each lead 
$\alpha=L$ and $R$ can be `split' into source ($s$) and drain ($d$), 
allowing the conductance through dot `1' (or `$N_c$') to be measured; see
Fig.\ref{cond schematic}. The Hamiltonian we study is
$\mathcal{H}^{N_c}=\mathcal{H}_L+\mathcal{H}^{N_c}_c+\mathcal{H}^{N_c}_{\text{hyb}}$, where the four equivalent
non-interacting leads are given by
\begin{equation}
\label{leads GV}
\mathcal{H}_L=\sum_{\substack{\alpha=L/R\\ \gamma=s/d}}\sum_{\textbf{k},\sigma}\epsilon^{\phantom{\dagger}}_{\textbf{k}}
c^{\dagger}_{\alpha\gamma \textbf{k} \sigma}c^{\phantom{\dagger}}_{\alpha\gamma
  \textbf{k} \sigma},
\end{equation}
and the impurity chain is described by
\begin{equation}
\label{chain GV}
\begin{split}
\mathcal{H}^{N_c}_{\text{c}} = & J \sum_{i=1}^{N_c-1}\hat{\textbf{S}}_{i} \cdot
\hat{\textbf{S}}_{i+1} \\
& +U(\hat{n}_{1\uparrow}\hat{n}_{1\downarrow} + \hat{n}_{N_c\uparrow}\hat{n}_{N_c\downarrow})+ \epsilon(\hat{n}_1 + \hat{n}_{N_c}), \\
\end{split}
\end{equation}
where
$\hat{n}_i=\sum_{\sigma}\hat{n}^{\phantom{\dagger}}_{i\sigma}=\sum_{\sigma}d^{\dagger}_{i
  \sigma}d^{\phantom{\dagger}}_{i \sigma}$ is the number operator for
dot $i=1$ or $N_c$, $\epsilon$ is its level energy and $U$ its Coulomb
repulsion/charging energy. In a real quantum dot device, the level energy is proportional to the gate voltage,
$\epsilon\propto V_g$. The leads and chain are coupled via
\begin{equation}
\label{hyb GV}
\mathcal{H}^{N_c}_{\text{hyb}} = 
\sum_{\substack{\gamma=s/d \\  \textbf{k},\sigma}}\left ( V^{\phantom{\dagger}}_{L\gamma}c_{L\gamma\textbf{k}\sigma}^{\dagger}d_{1\sigma}^{\phantom{\dagger}}
+
V^{\phantom{\dagger}}_{R\gamma}c_{R\gamma\textbf{k}\sigma}^{\dagger}d_{N_c\sigma}^{\phantom{\dagger}}
+ \text{H.c.} \right )
\end{equation}
where $V_{\alpha\gamma}$ is the tunnel-coupling matrix element for the
$\alpha=L/R$ and $\gamma=s/d$ lead. The hybridization strength follows
as $\Gamma_{\alpha\gamma}=\pi\rho_T V^2_{\alpha\gamma}$ (with
$\rho_T=N\rho$ the total lead density of states as before). Finally, a simple canonical
transformation of the lead orbitals, via
\begin{equation}
\label{rotation}
\begin{split}
c_{\alpha s \textbf{k}\sigma}=&a_{\alpha\textbf{k}\sigma}\cos(\theta_{\alpha})+\tilde{a}_{\alpha\textbf{k}\sigma}\sin(\theta_{\alpha})\\
c_{\alpha d \textbf{k}\sigma}=&a_{\alpha\textbf{k}\sigma}\sin(\theta_{\alpha})-\tilde{a}_{\alpha\textbf{k}\sigma}\cos(\theta_{\alpha})
\end{split}
\end{equation}
with $\tan(\theta_{\alpha})= V_{\alpha d}/V_{\alpha s}$, yields
an effective \emph{two}-channel model with
\begin{subequations}
\label{H GV 2leads}
\begin{align}
\label{2leads}
H_L&=\sum_{\alpha, \textbf{k},\sigma}\epsilon^{\phantom{\dagger}}_{\textbf{k}}
a^{\dagger}_{\alpha \textbf{k} \sigma}a^{\phantom{\dagger}}_{\alpha \textbf{k} \sigma}\\
\label{hyb 2leads}
\mathcal{H}^{N_c}_{\text{hyb}} &= 
\sum_{\textbf{k},\sigma}\left( V^{\phantom{\dagger}}_{L}  a_{L\textbf{k}\sigma}^{\dagger}d_{1\sigma}^{\phantom{\dagger}}
+
V^{\phantom{\dagger}}_{R} a_{R\textbf{k}\sigma}^{\dagger}d_{N_c\sigma}^{\phantom{\dagger}}
+ \text{H.c.}\right)
\end{align}
\end{subequations}
where $V_{\alpha}^2=V_{\alpha s}^2 + V_{\alpha d}^2$; so that in
particular for $V_{L}^2=V_{R}^2\equiv V^2$ (and hence $\Gamma_L = \Gamma_R
\equiv \Gamma=\pi\rho_T V^2$) the model is mirror-symmetric. 
To investigate 2CK physics on the lowest energy
scales, this is the situation now considered. We also focus on the
simplest example of the $N_c=3$ trimer, variants of which have been
studied recently in certain parameter regimes,\cite{ferrodots,zitkoTQD,akm_3dots,OguriTQD2009,TQD:FLNFL_zitko,zitkoTQD2ch,Martins2009,akm_frust}
including exchange-coupled chain\cite{TQD:FLNFL_zitko} and
ring\cite{zitkoTQD2ch,akm_frust} structures at half filling.
As shown below, a physically intuitive perturbative treatment of the model for different fillings
yield effective 2CK models -- in spin and orbital sectors -- from which the gate voltage-dependence of
the 2CK scale can be identified.


\subsection{Effective low-energy models for $N_c=3$}\label{GateV_eff}
In the atomic limit ($V=0$) of the isolated $N_c=3$ trimer, the number of chain
electrons jumps discontinuously between integer values
$\mathcal{N}=n_1+n_3+1=1\rightarrow 5$ as the gate voltage
$V_g\propto \epsilon$ is varied (recall that impurity `2'
is a strict spin-$\tfrac{1}{2}$). On tunnel-coupling to the
leads, this Coulomb blockade (CB) staircase is naturally smoothed
into a continuous crossover. Regimes of occupancy can still however be identified,
and sufficiently deep within the CB valleys, $\mathcal{N}$ will be approximately integral.
Here, a full $\mathcal{O}(V^2)$ Schrieffer-Wolff (SW) transformation\cite{hewson,SW}
can be performed in the strongly correlated regime of interest $U\gg
V$, perturbatively eliminating virtual excitations into 
high-energy manifolds with $(\mathcal{N}\pm 1)$ chain electrons. 

In the atomic limit the ground state in any given
$\mathcal{N}$-electron sector is a doublet, which we denote
$|\mathcal{N};\pm\tfrac{1}{2} \rangle$. Projecting onto the reduced (chain)
Hilbert space of this doublet using the unity operator
\begin{equation}
\label{unity n}
\hat{1}_{\mathcal{N}} = \sum_{\gamma=\pm\tfrac{1}{2}} |\mathcal{N};\gamma \rangle
\langle \mathcal{N};\gamma |,
\end{equation}
yields an effective model $\mathcal{H}_{\mathcal{N}}^{\text{eff}} = H_L +
\mathcal{H}^{\text{eff}}_{\mathcal{N},\text{II}}$, where the leading $\mathcal{O}(V^2)$
contribution arising from tunnel coupling to the leads 
(Eq.~\ref{hyb 2leads} with $V_L=V_R$ and $N_c=3$, denoted as
$\mathcal{H}'$) is given by the SW transformation\cite{hewson,SW} 
\begin{equation}
\label{SW}
\mathcal{H}^{\text{eff}}_{\mathcal{N},\text{II}} = \hat{1}_{\mathcal{N}}\mathcal{H}'(E_0-\mathcal{H}_c^{N_c=3})^{-1}\mathcal{H}'\hat{1}_{\mathcal{N}}.
\end{equation}
Here $E_0=\hat{1}_{\mathcal{N}} \mathcal{H}_c^{N_c=3} \hat{1}_{\mathcal{N}}$ is
the energy of the ground chain doublet (and  
retardation has as usual been neglected\cite{hewson}).

In the following, we also exploit the particle-hole transformation 
\begin{equation}
\label{ph}
\begin{split}
d_{i \sigma}^{\phantom{\dagger}} &\rightarrow d_{i \sigma}^{\dagger}, \qquad
~a_{\alpha\textbf{k}\sigma}^{\phantom{\dagger}} \rightarrow -a_{\alpha
  -\textbf{k}\sigma}^{\dagger}\\
\hat{S}_2^{\pm} &\rightarrow -\hat{S}_2^{\mp}, \qquad ~\hat{S}_2^z
\rightarrow -\hat{S}_2^z,
\end{split}
\end{equation}
which yields directly $\hat{\mathcal{N}}\rightarrow  (6-\hat{\mathcal{N}})$. 
The full Hamiltonian (parameterized by $\mathcal{H}\equiv
\mathcal{H}(\epsilon)$ for given $U$, $J$, $V$) 
transforms as $\mathcal{H}(\epsilon)\rightarrow \mathcal{H}(-\epsilon -U)
+(2U-4\epsilon)$. In general, the physical behavior of 
$\mathcal{H}(\epsilon)$ and $\mathcal{H}(-\epsilon-U)$ 
is equivalent, since the constant shift $(2U-4\epsilon)$ is irrelevant
in the calculation of observable quantities. 
Thus, the $\mathcal{N}=1\leftrightarrow 5$ and
$\mathcal{N}=2\leftrightarrow 4$ sectors are related by the
reflection about the particle-hole symmetric point $\epsilon =
-\tfrac{1}{2}U$. 
Together with the singly-occupied $\mathcal{N}=3$-electron
case, three distinct regions of electron filling must in consequence
arise. We now considered them in turn.


\subsubsection{$\mathcal{N}=3$ - electron regime}\label{GateV_3e}
For $-(U+\tfrac{1}{4}J)<\epsilon<\tfrac{1}{4}J$, the atomic limit trimer ground state
is a singly-occupied spin doublet\cite{akm_frust}
\begin{equation}
\label{state n=3}
\begin{split}
|\mathcal{N}=3;S^z \rangle = \tfrac{\sigma}{\sqrt{6}} \Big [ \hat{S}_{2}^{\sigma}
  (d_{1 \uparrow}^{\dagger} d_{3 \downarrow}^{\dagger} - &d_{3
    \uparrow}^{\dagger} d_{1 \downarrow}^{\dagger} ) \\
- & 2\hat{S}_{2}^{-\sigma} d_{1 \sigma}^{\dagger} d_{3 \sigma}^{\dagger}
\Big ] |\text{vac} \rangle,
\end{split}
\end{equation}
where $S^z=\tfrac{\sigma}{2}$ with $\sigma = \pm$ for spins
$\uparrow$$/$$\downarrow$, and $\hat{S}_{2}^{\sigma}\equiv
\hat{S}_{2}^{\pm}$ is a spin raising/lowering operator.
$|\text{vac}\rangle=\sum_{\sigma_2}|-;\sigma_2;-\rangle$ defines
the `vacuum' state of the local (chain) Hilbert space, in which dots `1'
and `3' are unoccupied, while `2' carries a free
spin-$\tfrac{1}{2}$. 

Using Eq.~\ref{state n=3} with Eqs.~\ref{unity n},\ref{SW} leads eventually to the
effective low energy  model deep in the $\mathcal{N}=3$ CB valley:
\begin{equation}
\label{Heff n=3}
\mathcal{H}^{\text{eff}}_{\mathcal{N}=3}=H_L+J_{K,\mathcal{N}=3}^{\text{eff}}
\hat{\textbf{S}}\cdot(\hat{\textbf{s}}_{L0}+\hat{\textbf{s}}_{R0})
\end{equation}
where we have omitted potential scattering contributions for clarity,
and $\hat{\textbf{S}}$ is a spin-$\tfrac{1}{2}$ operator for the
lowest chain doublet, defined by $\hat{S}^z =
\sum_{S^z} |\mathcal{N}=3;S^z\rangle
S^z \langle \mathcal{N}=3;S^z|$ and $\hat{S}^{\pm} = |\mathcal{N}=3;\pm\tfrac{1}{2}\rangle
\langle \mathcal{N}=3;\mp\tfrac{1}{2}|$. Eq.~\ref{Heff n=3} is 
of 2CK form, with effective Kondo coupling
\begin{equation}
\label{eff coup n=3}
\begin{split}
\rho J_{K,\mathcal{N}=3}^{\text{eff}}=&\frac{4\Gamma}{6\pi}\bigg{\lbrace} \frac{9}{J+4\epsilon+4U}+\frac{9}{J-4\epsilon}\\
&\quad-\frac{1}{5J+4\epsilon+4U}-\frac{1}{5J-4\epsilon}\bigg{\rbrace},
\end{split}
\end{equation}
which is AF throughout the entire $\mathcal{N}\approx 3$ sector. 
In the particle-hole symmetric Kondo limit in particular ($\epsilon=-\tfrac{1}{2}U$),
one obtains $\rho
J_{K,\mathcal{N}=3}^{\text{eff}}=\tfrac{2}{3}\rho J_K$ to leading
order in $1/U$ (with $\rho J_K = 8\Gamma/(\pi U)$ the effective
exchange coupling of a \emph{single} Anderson impurity\cite{hewson}
tunnel-coupled to leads); which as 
such is consistent with Eq.~\ref{eff coup bigj} and
Table~\ref{table:coup} for the $N_c=3$
Heisenberg chain studied in Sec.~\ref{largeJ}.

Two-channel Kondo physics thus arises in the $\mathcal{N}\approx 3$ regime, 
with $T_K^{2CK}$ in particular given from Eqs.~\ref{2CK pms}, \ref{eff coup n=3}.


\subsubsection{$\mathcal{N}=2,4$ - electron regime: orbital 2CK effect}\label{GateV_4e}
As above, the $\mathcal{N}=2$ and $4$ electron regimes are
related by the particle-hole transformation Eq.~\ref{ph}, so we consider
explicitly only the $\mathcal{N}=4$ case.  The $\mathcal{N}=4$ regime is the ground state
of the free trimer over an $\epsilon$-interval of width $J/2$, specifically
$-(U+\tfrac{3}{4}J ) <\epsilon<-(U+\tfrac{1}{4}J)$. The ground state comprises a degenerate
pair of spin singlets, since the spin-$\tfrac{1}{2}$ on 
impurity `2' can form a local singlet with either `1' or `3'
(the remaining site being doubly-occupied). Since the states are spin singlets, 
two-channel \emph{spin}-Kondo physics will obviously not arise here. 

 The $\mathcal{N}=4$ states are however doubly degenerate, so may be associated with an orbital pseudospin
($\hat{\boldsymbol{\mathcal{T}}}$), and expressed as
\begin{equation}
\label{state n=4}
\begin{split}
|\mathcal{N}=4;\mathcal{T}^z\rangle = \tfrac{\sigma}{\sqrt{2}}  \Big[ \Big(\hat{S}_{2}^{-}&
  d_{(2+\sigma) \uparrow}^{\dagger} 
  -\hat{S}_{2}^{+} d_{(2+\sigma)\downarrow}^{\dagger}\Big)\\
  & d_{(2-\sigma)
    \uparrow}^{\dagger} d_{(2-\sigma) \downarrow}^{\dagger})
\Big ] |\text{vac} \rangle,
\end{split}
\end{equation}
with $\mathcal{T}^z=\tfrac{\sigma}{2}$ for $\sigma=\pm 1$.
Projecting
into this reduced Hilbert space using the $\mathcal{N}=4$ unity operator
Eq.~\ref{unity n} with the SW transformation
Eq.~\ref{SW}, yields an effective \emph{orbital} 2CK model 
\begin{equation}
\label{Heff n=4}
\mathcal{H}^{\text{eff}}_{\mathcal{N}=4}=H_L+J_{K,\mathcal{N}=4}^{\text{eff}}
\hat{\boldsymbol{\mathcal{T}}} \cdot(\hat{\boldsymbol{\tau}}_{0,\uparrow}+\hat{\boldsymbol{\tau}}_{0,\downarrow}),
\end{equation}
with an effective exchange coupling
\begin{equation}
\label{eff coup n=4}
\begin{split}
\rho J_{K,\mathcal{N}=4}^{\text{eff}} =
\frac{\Gamma}{2\pi}\bigg{\lbrace}&
\frac{1}{\epsilon+U-\tfrac{3}{4}J}\\
&~~+\frac{2}{\epsilon+U+\tfrac{3}{4}J} -\frac{3}{\epsilon+U+\tfrac{1}{4}J}\bigg{\rbrace}
\end{split}
\end{equation}
which is AF within the $\mathcal{N}\approx 4$ regime. 
The trimer orbital pseudospin $\hat{\boldsymbol{\mathcal{T}}}$ is a
spin-$\tfrac{1}{2}$ operator defined by $\hat{\mathcal{T}}^z=\sum_{\mathcal{T}^z} |\mathcal{N}=4;\mathcal{T}^z\rangle
\mathcal{T}^z \langle \mathcal{N}=4;\mathcal{T}^z|$ and $\hat{\mathcal{T}}^{\pm} = |\mathcal{N}=4;\pm\tfrac{1}{2}\rangle
\langle \mathcal{N}=4;\mp\tfrac{1}{2}|$. Similarly, we may define a lead
pseudospin $\hat{\boldsymbol{\tau}}_{0,\sigma}$ for each real spin
$\sigma=\uparrow / \downarrow$ as
$\hat{\tau}_{0,\sigma}^z=\tfrac{1}{2}(f_{L0\sigma}^{\dagger}f_{L0\sigma}^{\phantom{\dagger}}-f_{R0\sigma}^{\dagger}f_{R0\sigma}^{\phantom{\dagger}})$ and
  $\hat{\tau}_{0,\sigma}^{+} =
    f_{L0\sigma}^{\dagger}f_{R0\sigma}^{\phantom{\dagger}}$ (with $\hat{\tau}_{0,\sigma}^{-} =
(\hat{\tau}_{0,\sigma}^{+})^{\dagger}$), where 
$f_{\alpha 0\sigma}^{\dagger}$ is given by Eq.~\ref{0 orb f}.

The important `pseudospin-flip' processes embodied in Eq.~\ref{Heff n=4}
correspond physically to moving an electron of given real spin from one
lead to the other, while simultaneously switching the trimer orbital
participating in the local singlet, such that no net charge transfer 
occurs between leads. 
Real spin $\sigma=\uparrow / \downarrow$ here plays the role of the channel
index, and as such orbital 2CK physics is expected below $T\sim T_K^{2CK}$, as
given by Eq.~\ref{2CK pms} using the effective coupling 
Eq.~\ref{eff coup n=4}.


\subsubsection{$\mathcal{N}=1,5$ - electron regime}\label{GateV_5e}

For $\epsilon < -(U+\tfrac{3}{4}J)$, the ground state in the atomic limit lies in the $\mathcal{N}=5$-electron regime. It is a spin doublet, comprising the free spin-$1/2$ on `impurity' 2, with sites `1' and `3' each doubly occupied: $|\mathcal{N}=5;S^z\rangle =   \hat{S}_{2}^{\sigma}
  d_{1 \uparrow}^{\dagger} d_{1 \downarrow}^{\dagger}  
  d_{3 \uparrow}^{\dagger} d_{3 \downarrow}^{\dagger}   
 |\text{vac} \rangle$.
Virtual excitations to the $\mathcal{N}=4$-electron sectors are perturbatively
eliminated to $\mathcal{O}(V^2)$ by the SW transformation
(Eq.~\ref{SW}), leading to
\begin{equation}
\label{Heff n=5}
\mathcal{H}^{\text{eff}}_{\mathcal{N}=5}=H_L+J_{K,\mathcal{N}=5}^{\text{eff}}
\hat{\textbf{S}}_2 \cdot(\hat{\textbf{s}}_{L0}+\hat{\textbf{s}}_{R0}),
\end{equation}
with effective coupling 
\begin{equation}
\label{eff coup n=5}
\rho J_{K,\mathcal{N}=5}^{\text{eff}} =
\frac{\Gamma}{\pi}\bigg{\lbrace}
\frac{1}{\epsilon+U-\tfrac{1}{4}J}
-\frac{1}{\epsilon+U+\tfrac{3}{4}J} \bigg{\rbrace}
\end{equation}
which is AF throughout the $\mathcal{N}\approx 5$-electron
regime. Two-channel spin-Kondo 
physics is thus again expected below $T\sim T_K^{2CK}$ (given via
Eq.~\ref{2CK pms}).


\begin{figure}
\includegraphics[height=6cm]{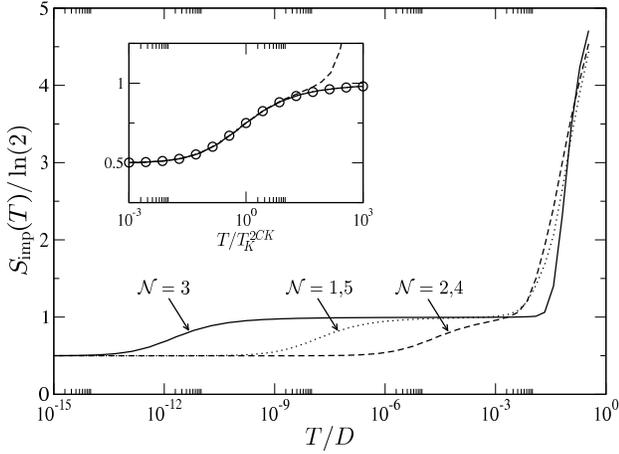}
\caption{\label{thermo gv}
Entropy $S_{\text{imp}}(T)$ \emph{vs} $T/D$ for a trimer with tunnel-coupled  
leads. Shown for $\rho J=0.075$, $U/\pi\Gamma=10$, $\Gamma/D=10^{-2}$,
varying $\epsilon/\pi\Gamma=-5, -12.5, -14.5$ (solid, dashed and
dotted lines), respectively representing systems `deep' in the $\mathcal{N}=3,4,5$ CB valleys. 
Identical results are obtained for their particle-hole transformed counterparts. Inset: the
low-temperature scaling behavior in terms of $T/T_K^{2CK}$, compared
with the standard 2CK model (circles).
}\end{figure}

\subsection{Thermodynamics and Scaling}\label{GateV_td}
The physical picture is clear: 2CK physics, whether of spin- or orbital-character,  is
expected when sufficiently deep in each region of electron filling. We confirm 
this directly using NRG for the full trimer model in Fig.~\ref{thermo
 gv}, where the entropy 
$S_{\text{imp}}(T)/\ln(2)$ \emph{vs} $T/D$ is shown for fixed representative
$\rho J$, $U/\pi\Gamma$, and $\Gamma/D$,
varying $\epsilon/\pi\Gamma$ for systems deep in the $\mathcal{N}=3,4,5$-electron CB valleys.

In each case, the high temperature $T>U$ behavior is simply that of
two free orbitals and a free spin, giving  $S_{\text{imp}}=5\ln(2)$. The LM FP 
is reached directly as $T$ is lowered, yielding $S_{\text{imp}}=\ln(2)$; 
flow to the 2CK FP with characteristic\cite{2ck:rev_cox_zaw}
$S_{\text{imp}}=\tfrac{1}{2}\ln(2)$ follows below $T\sim
T_K^{2CK}$. Upon rescaling in terms of $T/T_K^{2CK}$ (see inset), the systems in
each regime of filling collapse to the universal 2CK curve (circles),
thus confirming the effective low-energy models Eqs.~\ref{Heff n=3}, \ref{Heff n=4} and \ref{Heff n=5}.

\begin{figure}
\includegraphics[height=6cm]{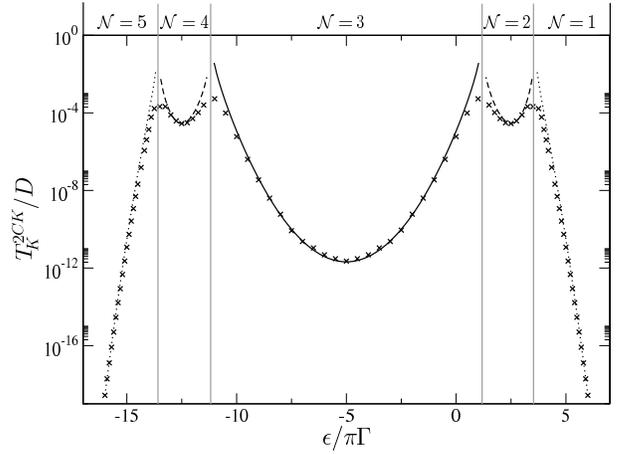}
\caption{\label{tk gv}
Evolution of the two-channel Kondo scale $T_K^{2CK}/D$ \emph{vs} $\epsilon/\pi\Gamma$
for $\rho J=0.075$, $U/\pi\Gamma=10$ and $\Gamma/D=10^{-2}$. 
NRG results (points) are compared with
Eq.~\ref{2CK pms} using effective Kondo couplings valid in the
$\mathcal{N}=3$ regime (Eq.~\ref{eff coup n=3}, solid line),
$\mathcal{N}=2,4$ regimes (Eq.~\ref{eff coup n=4}, dashed lines) and
the $\mathcal{N}=1,5$ regimes (Eq.~\ref{eff coup n=5}, dotted lines).
}\end{figure}

Fig.~\ref{tk gv} shows the evolution of the 2CK scale as the 
level energy ($\epsilon \propto V_g$) is varied essentially continuously over a wide range of
$\epsilon/\pi\Gamma$, for systems with the same 
$\rho J$, $U/\pi\Gamma$ and $\Gamma/D$ as in Fig.~\ref{thermo gv}. 
NRG results (points) are compared with the perturbative result for $T_K^{2CK}$ given in 
Eq.~\ref{2CK pms}, using the effective Kondo couplings valid in the
$\mathcal{N}=3$ regime (Eq.~\ref{eff coup n=3}),
$\mathcal{N}=2,4$ regimes (Eq.~\ref{eff coup n=4}) and
the $\mathcal{N}=1,5$ regimes (Eq.~\ref{eff coup n=5}). 
Throughout the majority of parameter space, the agreement is
excellent; only at the boundary between regimes does the perturbative
treatment (naturally) break down. Further, while the mechanism for overscreening changes
from spin-2CK (odd-$\mathcal{N}$) to orbital-2CK (even-$\mathcal{N}$) across these boundaries, 
$T_K^{2CK}$ itself is found from NRG to vary smoothly; the 2CK FP remaining the stable 
FP in all cases (including for $\epsilon/\pi\Gamma<-14$ and $>+4$ in Fig.~\ref{tk gv}, where $T_K^{2CK}$
diminishes rapidly but nonetheless remains finite).


\subsection{Single-particle dynamics and Conductance}\label{GateV_dyn}
We turn now to dynamics, focussing again on the spectrum
$-\pi\rho_T\text{Im}[t_{L}(\omega)]\equiv \pi\Gamma D_1(\omega)$,
where $D_1(\omega)=-\tfrac{1}{\pi}\text{Im}[G_{1}(\omega)]$ with
$G_{1}$ the local retarded Green function for dot `1'. We
obtain it through the Dyson equation,
\begin{equation}
\label{dyson}
[G_1(\omega)]^{-1}=[G_1^0(\omega)]^{-1}-\Sigma_1(\omega),
\end{equation}
where $G_1^0(\omega)$ is the non-interacting propagator (obtained for
$U=0=J$), and $\Sigma_1(\omega) =\Sigma_1^R(\omega) -
i\Sigma_1^I(\omega)$ is the proper electron self-energy. 
The non-interacting $G_1^0(\omega)$ is simply\cite{hewson}
$[G_1^0(\omega)]^{-1}=\omega^{+}-\epsilon-\Gamma(\omega)$ (with
$\omega^{+}=\omega+i0^{+}$); where $\Gamma (\omega) = \Gamma^{R}(\omega) -i\Gamma^{I}(\omega)$, 
with $\Gamma^{I}(\omega) = \Gamma$ (=$\pi V^{2}\rho_{T}$) for
all $|\omega| <D$ inside the band, and $\Gamma^{R}(\omega =0)=0$.

An expression for $\Sigma_{1}(\omega)$ is readily obtained using equation of motion methods,\cite{hewson,EOM} 
and is given by
\begin{equation}
\label{self-energy}
\begin{split}
\Sigma_{1}(\omega)=[G_{1}(\omega)]^{-1}\bigg{\{}&U\langle\langle
d^{\phantom{\dagger}}_{1\uparrow}\hat{n}^{\phantom{\dagger}}_{1\downarrow};d^{\dagger}_{1\uparrow}
\rangle\rangle_{\omega}^{\phantom\dagger}\\
+&\tfrac{1}{2}J\langle\langle
d^{\phantom{\dagger}}_{1\downarrow}
\hat{S}_2^{-}+d^{\phantom{\dagger}}_{1\uparrow}
\hat{S}_2^{z};d^{\dagger}_{1\uparrow}
\rangle\rangle_{\omega}^{\phantom\dagger} \bigg{\}}
\end{split}
\end{equation}
where the local Green function itself is 
$G_{1}(\omega)=\langle\langle
d^{\phantom{\dagger}}_{1\sigma};d^{\dagger}_{1\sigma}
\rangle\rangle_{\omega}^{\phantom\dagger}$ (independent of spin
$\sigma$ in the absence of a magnetic field, and 
with $G_{1}(\omega)=G_{3}(\omega)$ in the mirror-symmetric systems considered). The
self-energy can be calculated directly within the density matrix
formulation of NRG\cite{asbasis,fdmnrg,UFG,nrg_rev} via
Eq.~\ref{self-energy}; with $G_{1}(\omega)$ then obtained from
Eq.~\ref{dyson}.  In particular, the local propagator for $\omega=0$ may be expressed
simply as $[G_1(0)]^{-1}=-\epsilon^{*}+i\Gamma^{*}$ in
terms of the renormalized single-particle level $\epsilon^{*}$ and
renormalized hybridization $\Gamma^{*}$, given by
\begin{subequations}
\label{renorm}
\begin{align}
\epsilon^{*}&=\epsilon+\Sigma_1^R(0)\\
\Gamma^{*}&=\Gamma+\Sigma_1^I(0)
\end{align}
\end{subequations}
in terms of the self-energy at $\omega=0$. The Fermi level
value of the single-particle spectrum then follows as
\begin{equation}
\label{fermi level spec}
\pi\Gamma D_{1}(\omega=0)~=~\frac{^{\phantom *}\Gamma^{\phantom *}}{^{\phantom *}\Gamma^{*}}~\frac{1}{1+(\epsilon^{*}/\Gamma^{*})^{2}}. 
\end{equation}

\begin{figure}
\includegraphics[height=7cm]{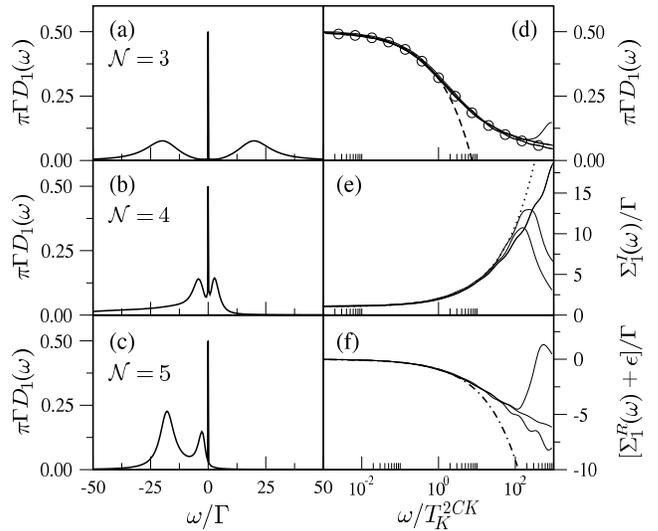}
\caption{\label{spec fig gv}
Left panels (a--c): $T=0$ single-particle spectrum $\pi\Gamma
D_1(\omega)$ \emph{vs} $\omega/\Gamma$. Shown for $\rho J=0.075$,
$U/\pi\Gamma=10$, $\Gamma/D=10^{-2}$, with  
$\epsilon/\pi\Gamma=-5$ [$\mathcal{N}=3$, panel (a)], $-12.5$
[$\mathcal{N}\approx 4$, panel (b)] and $-14.5$ 
[$\mathcal{N}\approx 5$, panel (c)]. Panel (d): collapse to
the universal 2CK scaling spectrum (circles). The
scaling behavior of the proper self-energy, plotted as 
$\Sigma_1^I(\omega)/\Gamma$ and
$[\Sigma_1^R(\omega)+\epsilon]/\Gamma$, is shown in panels (e) and
(f). Asymptotic $|\omega|\ll T_K^{2CK}$ behavior described by
$\pi\Gamma D_{1}(\omega)=\tfrac{1}{2}[1-b(|\omega|/T^{2CK}_K)^{1/2}]$ 
(dashed line); $\Sigma^I_{1}(\omega)/\Gamma
=1+2b(|\omega|/T_K^{2CK})^{1/2}$ (dotted); and 
$[\Sigma^R_{1}(\omega)+\epsilon]/\Gamma=-\text{sgn}(\omega)2b(|\omega|/T^{2CK}_K)^{1/2}$
(dot-dashed). 
}\end{figure}

NRG results are considered in Fig.~\ref{spec fig gv}, panels (a)--(c) showing $\pi\Gamma D_1(\omega)$
\emph{vs} $\omega/\Gamma$ for fillings $\mathcal{N}=3,4,5$, with the same parameters as used in Fig.~\ref{thermo gv}.
First, we comment briefly on the high-energy `Hubbard satellites' clearly visible in $D_1(\omega)$.
As usual,\cite{hewson} these reflect simple one-electron addition ($\omega>0$) or subtraction ($\omega<0$) from
the isolated chain $\mathcal{N}$-electron ground states. Their locations, broadened somewhat on coupling to the leads, are thus readily understood from the $V=0$ atomic limit ground states of Sec.~\ref{GateV_eff}.
Given their simplicity we do not comment further on them, save to note that in the $\mathcal{N}=3$ particle-hole symmetric example of panel (a), $D_1(\omega)=D_1(-\omega)$ as expected; and that for the $\mathcal{N}=5$
example in panel (c), high-energy features are naturally observed only for $\omega<0$, corresponding to excitations to $\mathcal{N}=4$-electron states.

The most important feature of $D_1(\omega)$ in Fig.~\ref{spec fig gv}
is of course the low-energy Kondo resonance, associated with RG flow
in the vicinity of the stable 2CK FP. This is shown in panel (d) where spectra from the
$\mathcal{N}=3,4,5$ regimes are again shown, but now rescaled in terms
of $\omega/T_K^{2CK}$. Collapse to a single curve is seen, with the
value at the Fermi level in particular pinned to $\pi\Gamma
D_1(\omega=0)=\tfrac{1}{2}$, \emph{independent} of $\epsilon$. The
low-$\omega$ asymptotics of the scaling spectrum (dashed line) are found to be
\begin{equation}
\label{lowfreqD}
\pi\Gamma D_{1}(\omega)~\overset{\tfrac{|\omega|}{T^{2CK}_{K}}\ll 1}\sim ~
\tfrac{1}{2}[1-b\left(|\omega|/T^{2CK}_K\right)^{1/2}],
\end{equation}
as consistent with behavior near the 2CK FP discussed in connection
with a variety of related two-channel models (see e.g.\ Refs.~\onlinecite{2ck:CFT_affleck2,TQD:FLNFL_zitko,2ck:dyn_johannesson,2ck:dyn_bradley,2ck:toth,akm_frust,2ck:dyn_anders}). Indeed, comparison to the spectrum $D\rho_K(\omega)$ for the 2CK model
shows perfect agreement in the low-energy scaling regime. This behavior is in striking contrast 
to that arising\cite{hewson} in a FL phase for $|\omega|\ll T_K$: 
$\pi\Gamma D(\omega)=\sin^2(\delta)-a_1(\omega/T_K)-a_2(\omega/T_K)^2$, describing 
the approach to the Fermi level value, which itself depends on the
phase-shift, $\delta$. For the Anderson model,\cite{hewson} $\delta=\tfrac{\pi}{2}n_{\text{imp}}$
by the Friedel sum rule,\cite{hewson,lang} with $n_{\text{imp}}$ the `excess' charge in the system induced by
addition of the impurity. Thus, $D(\omega=0)$ depends on the dot filling -- and hence on the level energy $\epsilon$ -- in a regular FL. The situation is clearly quite different in the stable NFL phase obtained for the chain models studied in the present work; and we shall consider the analogue of the Friedel sum rule in Sec.~\ref{GateV_lutt} below.

Further insight is however gained from the electron self-energy 
itself, the imaginary and real parts of which are shown respectively
in panels (e) and (f) of Fig.~\ref{spec fig gv}. To  emphasise the low-energy scaling of
interest, we  show the results for $\mathcal{N}=3,4,5$ in terms
of $\omega/T_K^{2CK}$.  The common asymptotic form for 
$|\omega|\ll T_K^{2CK}$ is found to be 
\begin{subequations}
\label{lowfreqSE}
\begin{align}
\Sigma^I_{1}(\omega)/\Gamma&~
\sim ~
1+2b(|\omega|/T_K^{2CK})^{1/2},\\
\Sigma^R_{1}(\omega)/\Gamma&~
\sim ~
-\epsilon/\Gamma-\text{sgn}(\omega)2b(|\omega|/T^{2CK}_K)^{1/2}
\end{align}
\end{subequations}
(with $b\sim {\cal{O}}(1)$ precisely the same constant as in Eq.~\ref{lowfreqD}).
At the Fermi level in particular, $\Sigma^I_1(\omega=0)=\Gamma$, in
contrast to generic FL behavior $\Sigma^I(\omega=0)=0$. Indeed, extensive
examination of NRG results over the entire parameter space confirms
Eq.~\ref{lowfreqSE} generally -- for \emph{any} value of the bare level energy $\epsilon$, and for all interaction strengths $U/\Gamma$ and exchange couplings $\rho J >0$.
The renormalized level energy and hybridization then follow from Eq.~\ref{renorm} as $\epsilon^{*}=0$ and
$\Gamma^{*}=2\Gamma$. From Eq.~\ref{fermi level spec}, the spectrum at the Fermi level is in
consequence pinned to a universal half-unitarity value,
$\pi\Gamma D_1(\omega=0)=\tfrac{1}{2}$ for all underlying bare parameters, as
illustrated in panel (d) of Fig.~\ref{spec fig gv}.


\subsubsection{Conductance}\label{GateV_Gc}

To measure the differential conductance through dot `1', a bias voltage $V_{sd}$ is applied across the $L$ 
source and drain leads, inducing a chemical potential difference $\mu_s-\mu_d = eV_{sd}$. The $L/R$ symmetry required to observe 2CK physics on the lowest energy scales also requires of course that 
the same bias be applied across the $R$ lead.
Following Meir and Wingreen,\cite{wingreenspec}
the zero-bias conductance through dot `1' is given exactly by
\begin{equation}
\label{MW0}
G_c=\frac{2e^2}{h}G_0 \int_{-\infty}^{\infty} \text{d}\omega 
  (-)\frac{\partial f(\omega)}{\partial \omega}~ \pi \Gamma D_1(\omega) 
\end{equation}
where $D_{1}(\omega)$ is the single-particle spectrum
at equilibrium, $f(\omega)=[e^{\omega/T}+1]^{-1}$, and $\Gamma=\Gamma_s+\Gamma_d$
is the total hybridization as before. The dimensionless 
$G_0=4\Gamma_s\Gamma_d/(\Gamma_s+\Gamma_d)^2$ embodies simply the relative
coupling to source and drain leads; such that for $\Gamma_s =\Gamma_d$
$G_0=1$ is maximal, while in the extreme asymmetric limit $\Gamma_s\gg \Gamma_d$ (where the drain
acts as a weak tunneling probe), $G_0\sim 4\Gamma_d/\Gamma_s\ll 1$.

For $T=0$, Eq.~\ref{MW0} reduces simply to
\begin{equation}
\label{zbc}
G_c^0=\frac{2e^2}{h}G_0\pi\Gamma D_1(\omega=0),
\end{equation}
and Eq.~\ref{lowfreqD} then
gives a universal zero-bias conductance 
$G_c^0/G_0=e^2/h$ at $T=0$, obtained for any value of the gate voltage
$V_g\propto \epsilon$. This result is thus consistent with that known for
related models in the singly-occupied Kondo limit which demonstrate 2CK behavior 
(see e.g.\ Refs.~\onlinecite{2ck:proposal,2ck:expt_potok,2ck:cond_pustilnik,2ck:chains_zarand,TQD:FLNFL_zitko,2ck:dyn_toth,akm_frust}). For finite $T$, the Fermi level value of the 
spectrum has the same low-$T/T_{K}^{2CK}$ 
dependence\cite{akm_frust} as the $T=0$ spectrum does of $\omega/T_{K}^{2CK}$ (Eq.~\ref{lowfreqD}),  viz.\ 
$\pi\Gamma D_{1}(\omega =0; T) \sim \tfrac{1}{2}[1-b^{\prime}(T/T_{K}^{2CK})^{1/2}]$.
Combined with Eq.~\ref{lowfreqD}, Eq.~\ref{MW0} is then readily shown\cite{akm_frust} to yield
$G_{c}(T)/G_{0} \sim \tfrac{e^2}{h}[1-\gamma(T/T_{K}^{2CK})^{1/2}]$, 
with $\gamma =b^{\prime}+b\sqrt{\pi}\eta(\tfrac{1}{2})$ and
$\sqrt{\pi}\eta(\tfrac{1}{2}) \simeq 1.07$; 
which $T$-dependence is also known to arise for the single-spin 2CK model.~\cite{2ck:CFT_affleck2,2ck:cond_pustilnik,2ck:dyn_toth,2ck:toth}

 Calculating the conductance at finite bias is of course a different matter, and an 
exact (or even numerically exact) treatment of the underlying non-equilibrium
physics is a formidable open problem. Here we merely make the simplifying approximation
that the self-energy does not depend explicitly on the bias voltage,\cite{MRGnanoprl} which leads to
\begin{equation}
\label{MW}
G_c \simeq \frac{e^2}{h}G_0 \int_{-\infty}^{\infty} \text{d}\omega \left (
  -\frac{\partial f_s(\omega)}{\partial \omega}-\frac{\partial
    f_d(\omega)}{\partial \omega}\right ) \pi \Gamma D_1(\omega) 
\end{equation}
where $f_{\gamma}(\omega)=[e^{(\omega-\mu_{\gamma})/T}+1]^{-1}$ is the
Fermi function for the $\gamma=s/d$ lead.
While this approximation is exact both for $V_{sd}= 0$, and for all $V_{sd}$ in the  extreme asymmetric limit $\Gamma_s\gg \Gamma_d~$\cite{kondo:rev_pustilnik}  (conductance measured with a `perfect STM'),
in the standard case relevant to semiconductor quantum dot devices the leads are more symmetrically coupled. Here we consider a symmetric voltage split between the leads, $\mu_{s/d}=\pm \tfrac{1}{2}eV_{sd}$, for which Eq.~\ref{MW} yields
\begin{equation}
\label{finite V}
G_c(V_{sd})\simeq \frac{e^2}{h}G_0\pi\Gamma \left [ D_1(\omega=\tfrac{1}{2}
  eV_{sd}) + D_1(\omega=-\tfrac{1}{2} eV_{sd}) \right ]
\end{equation}
for $T=0$. This approximation thus allows us to work with single-particle spectra determined at equilibrium, 
obtained from a two-lead NRG calculation\cite{asbasis,fdmnrg,UFG,nrg_rev} as before.

\begin{figure}
\includegraphics[height=5cm]{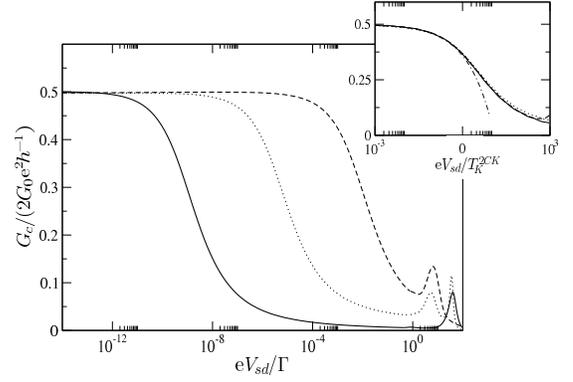}
\caption{\label{cond fig gv}
Conductance $G_c/(2G_0e^2h^{-1})$ \emph{vs} $eV_{sd}/\Gamma$ for systems with
same parameters as Figs.~\ref{thermo gv},\ref{spec fig gv}. Inset: scaling collapse to a universal curve with asymptotic
$eV_{sd}\ll T_K^{2CK}$ behavior $G_c(V_{sd})=G_0 e^2
h^{-1}[1-b(eV_{sd}/T_K^{2CK})^{1/2}]$ (dot-dashed line).
}\end{figure}

Fig.~\ref{cond fig gv} shows the resultant differential conductance
$G_c/(2G_0e^2h^{-1})$ \emph{vs} $eV_{sd}/\Gamma$, calculated using
Eq.~\ref{finite V}, for the same
$\mathcal{N}=3,4,5$ systems as in Figs.~\ref{thermo gv} and \ref{spec
  fig gv}. Since the conductance comprises a symmetrized combination
of the total dot spectrum, similar features to that in 
Fig.~\ref{spec fig gv} are naturally observed. Peaks at high bias 
originate from the Hubbard satellites, and correspond to simple
single-electron sequential tunelling processes. Importantly, the Kondo
resonance also of course shows up, with the zero-bias
value of $G_c^0/G_0=e^2/h$ arising for $eV_{sd}\ll T_K^{2CK}$
in each case. Universal scaling of the conductance in terms of
$eV_{sd}/T_K^{2CK}$ is also shown in the inset, demonstrating in
particular the  $eV_{sd}\ll T_K^{2CK}$ asymptotic form
\begin{equation}
\label{cond scale}
G_c(V_{sd}) ~\overset{\tfrac{eV_{sd}}{T^{2CK}_{K}}\ll 1}\sim ~ 
\frac{e^2}{h}G_0[1-b(eV_{sd}/T_K^{2CK})^{1/2}],
\end{equation}
which behavior is likewise
known\cite{2ck:proposal,2ck:expt_potok,2ck:cond_pustilnik} in the NFL
regime of the 2CK device constructed in Ref.~\onlinecite{2ck:expt_potok}.


\subsection{Phase shifts and the S-matrix}\label{Smatrix}

We now consider the leading low-$\omega$ behavior of the single-particle scattering S-matrix, 
$S(\omega) =e^{2i\delta(\omega)}$, and
associated phase shift $\delta(\omega)=\delta_R(\omega)+i\delta_I(\omega)$.
The S-matrix is given by\cite{lang}
\begin{subequations}
\label{smatrix}
\begin{equation}
\label{smatrix t}
S(\omega) ~=~ 1-2i\Gamma G_{1}(\omega),
\end{equation}
with $\Gamma G_{1}(\omega)$ related to the $t_{L}(\omega)$-matrix  by
\begin{equation}
\label{tG}
\pi\rho_{T} t_{L}(\omega) ~=~ \Gamma G_{1}(\omega)
\end{equation}
\end{subequations}
such that (as in Sec.~\ref{GateV_dyn}) $-\pi\rho_{T}\mathrm{Im}t_{L}(\omega)=\pi\Gamma D_{1}(\omega)$. It follows 
from Eq.~\ref{smatrix t} that
\begin{equation}
\label{t ps}
\pi\Gamma D_{1}(\omega) = \tfrac{1}{2}\left \{ 1-
  e^{-2\delta_I(\omega)}\cos (2\delta_R(\omega))\right \},
\end{equation}
leading in particular to the limiting behavior:
\begin{equation}
\label{t-matrix ps}
\pi\Gamma D_{1}(\omega) =
\begin{cases}
\sin^2(\delta_R(\omega)) &~~:\delta_I(\omega)\rightarrow 0\\
\tfrac{1}{2} &~~:\delta_I(\omega)\rightarrow \infty
\end{cases}
\end{equation}

To obtain $S(\omega)$ it is convenient to express the propagator as
$G_{1}(\omega) ~=~ [A(\omega)+iB(\omega)]^{-1}$, where
\begin{equation}
\label{AB}
A(\omega)~=~ \omega -\epsilon -\Sigma_{1}^{R}(\omega), ~~~ B(\omega)~=~ \Gamma + \Sigma_{1}^{I}(\omega)
\end{equation}
such that $A(0)=-\epsilon^{*}$ and $B(0)=\Gamma^{*}$ by Eq.~\ref{renorm}
(and for simplicity we take here the wide-band limit for the lead density of states,
$\Gamma(\omega) =-i\Gamma$, which does not affect any of the following results). From Eq.~\ref{smatrix t}
it follows that
\begin{equation}
\label{eid}
e^{2i\delta(\omega)}~=~ \frac{A(\omega)-iB^{\prime}(\omega)}{A(\omega)+iB(\omega)}
\end{equation}
where
\begin{equation}
\label{Bprime}
B^{\prime}(\omega)~=~ \Gamma - \Sigma^{I}_{1}(\omega).
\end{equation}
 
 First consider the familiar situation that would arise if the system was a 
regular Fermi liquid, for which $\Sigma_{1}^{I}(\omega =0) =0$. In this case, 
$B(0)=B^{\prime}(0)$, and Eq.~\ref{eid} yields $\delta (0) = \mathrm{arg}[G_{1}(0)]$.
The S-matrix is then unitary at the Fermi level, $|S(0)|^{2} =1$; and, since $\delta_{I}(0)=0$, the Fermi level spectrum follows from Eq.~\ref{t-matrix ps} as $\pi\Gamma D_{1}(0) = \mathrm{sin}^{2}(\delta_{R}(0))$.

The situation is of course quite different for the present problem. The 
low-frequency behavior of the self-energy is given by Eqs.~\ref{lowfreqSE},
and from which Eqs.~\ref{AB}-\ref{Bprime} yield
\begin{equation}
\label{eidlow}
e^{2i\delta(\omega)}~
\sim ~b \left(\frac{|\omega|}{T_{K}^{2CK}}\right)^{\frac{1}{2}}\left[ 1-i~\mathrm{sgn}(\omega)\right]
\end{equation}
as the leading asymptotic form for $|\omega|/T_{K}^{2CK} \rightarrow 0$; i.e.\
\begin{subequations}
\begin{equation}
\label{deltaR}
\delta_{R}(\omega)~\sim ~ \tfrac{1}{2}\mathrm{arctan}[-\mathrm{sgn}(\omega)]
\end{equation}
\begin{equation}
\label{deltaI}
e^{-2\delta_{I}(\omega)}~\sim~\sqrt{2}b \left(\frac{|\omega|}{T_{K}^{2CK}}\right)^{\frac{1}{2}}.
\end{equation}
\end{subequations}
In evident contrast to a FL, the imaginary part of the phase shift thus diverges
logarithmically as $\omega \rightarrow 0$,
\begin{equation}
\label{logdiv}
\delta_{I}(\omega)~\sim ~ -\frac{1}{4}\ln (|\omega|/T_{K}^{2CK}),
\end{equation}
the divergence itself 
reflecting (see Eq.~\ref{t-matrix ps}) the pinning
of the Fermi level spectrum to a half-unitary value
(Sec.~\ref{GateV_dyn}). In consequence, the S-matrix vanishes at the Fermi level, $S(0)=0$, 
as known for the single spin-$\tfrac{1}{2}$ 2CK model.~\cite{2ckMaldacena,2ckZarandscatt,2ck:scatter} 
This does not of course mean that an electron sent in to scatter off the dot is `absorbed' 
(the conductance being generically non-zero), but rather that electrons scatter completely into collective excitations, characteristic of the NFL state.~\cite{2ckMaldacena,2ck:scatter}

Notice also from Eq.~\ref{deltaR} that the real part of the phase shift is discontinuous across the Fermi 
level, and that $\delta_{R}(\omega =0\pm) \neq \mathrm{arg}[G_{1}(0)]$ (again in contrast to a FL).
$\mathrm{cos}(2\delta_{R}(0)) =1/\sqrt{2}$ is however continuous across $\omega =0$ and, combined with
Eq.~\ref{deltaI}, Eq.~\ref{t ps} recovers precisely Eq.~\ref{lowfreqD} for the low-$\omega$ asymptotics of
$\pi\Gamma D_{1}(\omega)$.

\subsection{Friedel-Luttinger sum rule}\label{GateV_lutt}
We now consider further implications of the pinning of 
the $T=0$ Fermi level spectrum, $\pi\Gamma D_{1}(0) =\tfrac{1}{2}$, regardless of 
bare model parameters and even when the dot occupancies change drastically on varying the 
bare level energy $\epsilon$. In particular, we obtain an analogue of the Friedel sum rule\cite{hewson,lang} -- a Friedel-Luttinger sum rule\cite{CJW_multilevel} -- relating the Fermi level spectrum to the `excess' charge 
induced by addition of the impurity chain,\cite{hewson} via the Luttinger 
integral.\cite{lutt,lutt_ward}

\begin{figure}
\includegraphics[height=6cm]{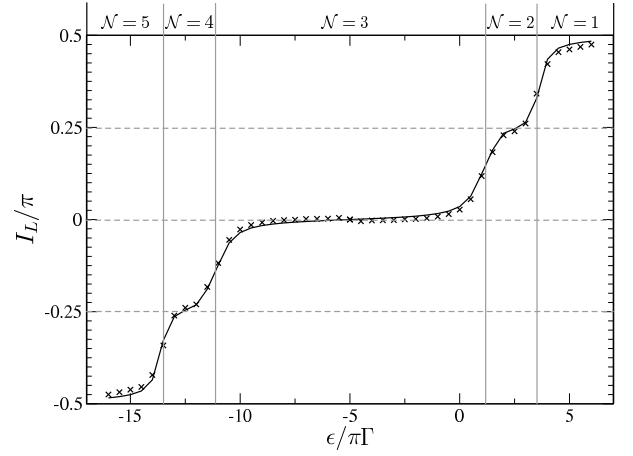}
\caption{\label{nrg lutt}
Luttinger integral $I_L/\pi$ \emph{vs} level energy
$\epsilon/\pi\Gamma$ for systems with $\rho J= 0.075$, $U/\pi\Gamma=10$
and $\Gamma/D=10^{-2}$. Direct calculation via Eq.~\ref{luttinger int}
shown as points, compared with Eq.~\ref{luttinger vary} (line) using
$n_{\text{imp}}$ determined from a standard thermodynamic NRG
calculation.
}\end{figure}

To this end, consider first the excess charge $n_{\mathrm{imp}}$, defined as the difference in charge of the entire system with and without the trimeric impurity chain;\cite{hewson} and also $n_{\mathrm{imp}}^{\prime}$, defined correspondingly but with only the two terminal dots (`1' and `3') of the chain removed. Since impurity `2' is a strict spin, it follows trivially that $n_{\mathrm{imp}}=n_{\mathrm{imp}}^{\prime}+1$. Using e.g.\ equation of motion methods,\cite{hewson,EOM} it is readily shown that 
\begin{equation}
\label{nimp}
n_{\text{imp}}^{\prime}=
-\frac{4}{\pi}\text{Im}\int_{-\infty}^{0}\text{d}\omega ~
G_1(\omega)\left [1- \frac{\partial\Gamma(\omega)}{\partial\omega}
\right ]
\end{equation}
(noting that sites `1' and `3' are equivalent by symmetry). In practice, as expected physically, $n_{\mathrm{imp}}^{\prime}$ differs negligibly from the charge $2\langle\hat{n}_{1}\rangle$ on the terminal dots, to which it reduces precisely in the wide flat-band limit where $\Gamma (\omega )=-i\Gamma$ is constant.

Note next that Eq.~\ref{fermi level spec} can be written as
\begin{equation}
\label{Dtheta}
\pi\Gamma D_1(\omega=0)=(\Gamma/\Gamma^{*}) \sin^2(\theta) 
\end{equation}
with $\theta =\mathrm{arctan}(\Gamma^{*}/\epsilon^{*})$ ($\equiv \mathrm{arg}[G_{1}(0)]$).
Equivalently, using $\mathrm{arg}[G_{1}(\omega =-\infty)] = 0$, 
\begin{equation}
\label{theta}
\theta=\text{Im}\int_{-\infty}^{0}\text{d}\omega
\frac{\partial}{\partial\omega}\ln G_1(\omega).
\end{equation}
But from the definition of the propagator, 
$G_1(\omega)=[\omega^{+}-\epsilon-\Gamma(\omega)-\Sigma_1(\omega)]^{-1}$, 
it follows that
\begin{equation}
\label{deriv}
\frac{\partial}{\partial \omega}\ln G_1(\omega) = -G_1(\omega)\left [
 1- \frac{\partial\Gamma(\omega)}{\partial \omega}\right ] +
G_1(\omega)\frac{\partial \Sigma_1(\omega)}{\partial\omega}.
\end{equation}
The Friedel-Luttinger sum rule then follows directly from Eq.~\ref{theta} as
\begin{equation}
\label{fsr}
\theta = \frac{\pi}{4} n_{\text{imp}}^{\prime} + I_L
\end{equation}
where the Luttinger integral~\cite{lutt,lutt_ward}
\begin{equation}
\label{luttinger int}
I_L=\text{Im}\int_{-\infty}^{0}\text{d}\omega ~
G_1(\omega)\frac{\partial\Sigma_1(\omega)}{\partial\omega}
\end{equation}
involves integration over all energy scales.

Again consider briefly the situation that would arise if the system were a normal FL. In this case 
the Luttinger integral vanishes\cite{lutt,lutt_ward} regardless of bare model parameters, and
$\theta \equiv \delta_{R}(0)$ (as in Sec.~\ref{Smatrix}). Eq.~\ref{fsr} then reduces to a Friedel sum rule,~\cite{hewson, lang} relating the static phase shift to the excess charge.

The present problem is not of course a Fermi liquid, and $I_{L}$ does not vanish.
The single-particle spectrum is however ubiquitously pinned at
$\pi\Gamma D_1(\omega=0)=\tfrac{1}{2}$  (i.e.\ 
$\Gamma^{*}=2\Gamma$ and $\epsilon^{*}=0$), 
whence $\theta=\pi/2$ regardless of bare parameters.

Eq.~\ref{fsr} in this case thus becomes a sum rule relating the Luttinger integral
to the excess charge:
\begin{equation}
\label{luttinger vary}
I_L~=~ \frac{\pi}{4}(2-n_{\text{imp}}^{\prime}) ~~ = \frac{\pi}{4}(3-n_{\text{imp}})
\end{equation}
Under the particle-hole transformation Eq.~\ref{ph} it is easily shown that
$n^{\prime}_{\mathrm{imp}} \rightarrow 4-n^{\prime}_{\mathrm{imp}}$
(or equivalently $n_{\mathrm{imp}}\rightarrow 6-n_{\mathrm{imp}}$). Hence $I_{L} \rightarrow -I_{L}$ 
under the transformation, and in particular vanishes at the particle-hole symmetric point $\epsilon = -U/2$
where $n_{\mathrm{imp}}=3$ identically.

The behavior of $n_{\mathrm{imp}}$ as e.g.\ the level energy $\epsilon$ is varied, is naturally a 
smoothed/continuous version of the Coulomb blockade staircase arising in the atomic limit (where on varying $\epsilon$ the total number of electrons in the free chain, ${\cal{N}}$, jumps discontinuously between integer values characteristic of each CB valley). $I_{L}$ itself will thus reflect that variation; and
sufficiently deep in each CB valley, where $n_{\mathrm{imp}}$ ($\simeq {\cal{N}}$) is close to integral, 
each regime may be loosely associated with its own value of $I_{L}$.

The above discussion is exemplified clearly by Fig.~\ref{nrg lutt}, where the Luttinger
integral $I_L/\pi$ is shown \emph{vs} the level energy
$\epsilon/\pi\Gamma$ for systems with common $U/\pi\Gamma=10$ and
$\rho J=0.075$. The points correspond to direct calculation of $I_{L}$ 
via Eq.~\ref{luttinger int}, using the full Green function and
self-energy from NRG. The line is simply Eq.~\ref{luttinger
 vary}, using $n_{\text{imp}}$ as determined from a standard
thermodynamic NRG calculation.\cite{KWW} The agreement is excellent over the
wide range of electron fillings shown, the overall form of the curve reflecting
the smoothed CB staircase as anticipated above. \\

 Our focus has been the $N_{c}=3$ trimer, but for longer (odd) chains one naturally expects the same 2CK physics to occur on low-energy scales. We have indeed confirmed this explicitly by NRG for the $N_{c}=5$ case. In particular, the single-particle spectrum of dot `1' at the Fermi level is again always pinned to half-unitarity, $\pi\Gamma D_1(\omega=0)=\tfrac{1}{2}$ (whence the zero-bias conductance through a terminal dot remains $G_c/G_0=e^2/h$). The result $\theta=\pi/2$ thus holds generally, as does the Friedel-Luttinger sum rule Eq.~\ref{fsr}, with $n_{\mathrm{imp}}^{\prime}$ now related to the total excess charge by $n_{\mathrm{imp}} = n_{\mathrm{imp}}^{\prime}+(N_{c}-2)$; and in consequence the general result for the Luttinger integral for odd chains follows:
\begin{equation}
\label{lutt gen}
I_L=~ \frac{\pi}{4}(2-n_{\text{imp}}^{\prime}) ~~ = \frac{\pi}{4}(N_c-n_{\text{imp}})
\end{equation}


\section{Concluding Remarks: Real Quantum Dot systems}\label{real dots}

The exchange-coupled impurity chains studied in this paper may be considered as approximate low-energy 
models of quantum dot devices. In real systems, however, the dots are mutually tunnel-coupled rather than pure exchange-coupled; the 2CK fixed point is rendered unstable by the inter-lead charge transfer 
that results, and the system crosses over to FL behavior on a low-energy scale 
$T \lesssim T_{\text{FL}}$. Experimental access to 2CK physics must thus contend with 
both channel anisotropy (as studied in Sec.~\ref{chains}), and charge transfer terms. On the level of a toy model calculation, we now consider briefly the generic behavior arising when the latter perturbation is included, considering explicitly the $L/R$ mirror symmetric case (although the analysis is readily extended to include explicit channel anisotropy).

 To motivate this, recall that a single ($N_{c}=1$) one-level quantum dot tunnel-coupled to two metallic leads does not of course exhibit 2CK physics.\cite{kondo:rev_pustilnik,1dot:Goldhaber,1dot:Cronenwett} 
This follows from the Anderson Hamiltonian itself,  
$H_{\text{And}}=H_{\text{dot}}+\sum_{\alpha=L/R}[H^{\alpha}_L+H^{\alpha}_{\text{hyb}}]$;
where $H_{\text{dot}} = \epsilon(\hat{n}_{d\uparrow}+\hat{n}_{d\downarrow})+ U\hat{n}_{d\uparrow}\hat{n}_{d\downarrow}$, 
$H^{\alpha}_L  = \sum_{\textbf{k},\sigma}\epsilon_{\textbf{k}}a_{\alpha\textbf{k}\sigma}^{\dagger}
a_{\alpha\textbf{k}\sigma}^{\phantom{\dagger}}$ and
$H^{\alpha}_{\text{hyb}}=
V\sum_{\textbf{k},\sigma}(a_{\alpha\textbf{k}\sigma}^{\dagger}d^{\phantom{\dagger}}_{\sigma}
+ \text{H.c.})$.
Transforming canonically to even ($e$) and odd ($o$) lead orbitals 
\begin{equation}
\label{evenodd}
\begin{split}
a_{e\textbf{k}\sigma} = &\tfrac{1}{\sqrt{2}} \left (
  a_{L\textbf{k}\sigma} + a_{R\textbf{k}\sigma} \right )\\
a_{o\textbf{k}\sigma} = &\tfrac{1}{\sqrt{2}} \left (
  a_{L\textbf{k}\sigma} - a_{R\textbf{k}\sigma} \right ),
\end{split}
\end{equation}
$H_{\text{And}}$ is equivalent to 
$H_{\text{And}}=H_{\text{dot}}+H^{e}_L+\sqrt{2}H^{e}_{\text{hyb}}$,
in which the dot couples solely to the $e$-lead; exhibiting as such 
single-channel physics only.

In the singly-occupied dot regime, a low-energy spin-$\tfrac{1}{2}$ Kondo 
model follows from a SW transformation\cite{hewson,SW} of $H_{\text{And}}$, leading 
simply to
$H_{\text{SW}}=H_L^{L}+H_L^{R} + H_K$ with
\begin{equation}
\label{kondo}
H_K= J_{K} \hat{\textbf{S}} \cdot (
\hat{\textbf{s}}_{L0} + \hat{\textbf{s}}_{R0} ) +
J_{LR}\hat{\textbf{S}} \cdot \hat{\textbf{s}}_{LR 0}
\end{equation}
(potential scattering is ignored); where $\hat{\textbf{s}}_{\alpha 0}$ is
given by Eq.~\ref{0 orb}, and $\hat{\textbf{s}}_{LR 0}$ is defined as
\begin{equation}
\label{JLR def}
\hat{\textbf{s}}_{LR 0}=\sum_{\sigma,\sigma',\alpha} f^{\dagger}_{\alpha 0 \sigma} \boldsymbol{\sigma}_{\sigma\sigma'} f^{\phantom{\dagger}}_{\bar{\alpha} 0 \sigma'}
\end{equation}
with $\bar{\alpha} = R,L$ for $\alpha =L,R$.
Eq.~\ref{kondo} consists formally of a symmetric 
2CK model -- the first term -- together with a term
$\hat{\textbf{S}} \cdot \hat{\textbf{s}}_{LR 0}$ which 
transfers (cotunnels) charge between the leads. In fact, applying the transformation
Eq.~\ref{evenodd} yields $H_{\text{SW}}=H_L^e+H_{L}^{o}+H_K$ with
\begin{equation}
\label{kondo asym}
H_K= J_{K} \hat{\textbf{S}} \cdot (
\hat{\textbf{s}}_{e0} + \hat{\textbf{s}}_{o0} ) +
J_{LR}\hat{\textbf{S}} \cdot (
\hat{\textbf{s}}_{e0} - \hat{\textbf{s}}_{o0} ),
\end{equation}
where $\hat{\textbf{s}}_{e0}$ and $\hat{\textbf{s}}_{o0}$ are $e/o$ lead spin densities; a model
that is generically of channel-asymmetric 2CK form. But for the single-dot Anderson model itself
the couplings are necessarily equal, $J_{LR}=J_K$. In this case the dot 
is exchange-coupled solely to the even lead, and Eq.~\ref{kondo asym}
reduces as it must to a single-channel Kondo model with Kondo coupling $2J_{K}$.

In systems comprising several tunnel-coupled quantum dots, however, 
cotunneling charge-transfer can be effectively suppressed,\cite{2ck:chains_zarand}  with $J_{LR} \ll J_{K}$ expected for longer 
chains (a simple estimate yielding $J_{LR}/J_K\sim [t/U]^{N_c-1}$ with
$t$ the inter-dot tunnel coupling). Here we simply regard
Eq.~\ref{kondo}, with $J_{LR}\neq J_{K}$, as an effective toy model to
mimic such effects in odd-$N_{c}$ dot chains (with $\hat{\textbf{S}}$
representing the lowest chain doublet). 
From Eq.~\ref{kondo asym} it is clear that the resultant low-energy/$T$ physics
is then that of the channel-asymmetric 2CK model.\cite{2ck:nozieres,2ck:rev_cox_zaw,2ck:BA_sacramento,2ck:BA_andrei_jerez,2ck:CFT_NRG} The 2CK FP is thus rendered unstable by the perturbation $J_{LR}$, any nascent 2CK state forming at $T_K^{2CK}$ being destroyed below the FL crossover scale $T_{\text{FL}}$; although inclusion of the $J_{LR}$ term 
should not obscure the 2CK physics for $T\gg T_{\text{FL}}$, provided $J_{LR}$ itself is
sufficiently small. Note also that any \emph{direct} inter-lead tunneling terms of the type 
$\sum_{\mathbf{k},\mathbf{k}^{\prime}}(a_{L\textbf{k}\sigma}^{\dagger}a^{\phantom{\dagger}}_{R\textbf{k}'\sigma} + \text{H.c.})$ -- as opposed to cotunneling which intrinsically proceeds \emph{via} the 
dot spin -- are equivalent (through the transformation Eq.~\ref{evenodd}) to simple 
potential scattering in the even and odd channels. This does not destabilise the 2CK FP,\cite{2ck:CFT_affleck1} 
for which reason we do not include such terms here.

\begin{figure}
\includegraphics[height=8cm]{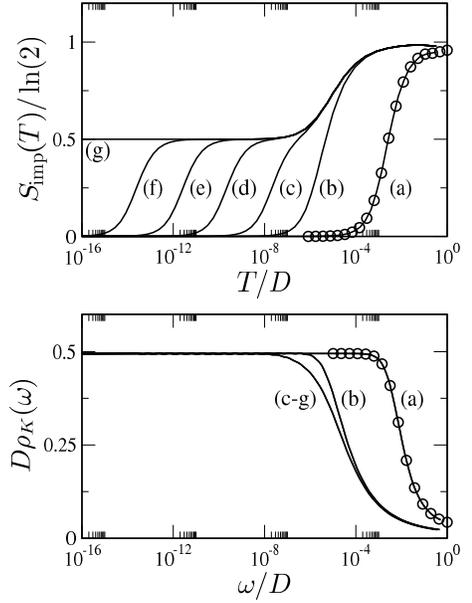}
\caption{\label{jlr pert}
Upper panel: entropy $S_{\text{imp}}(T)$ \emph{vs} $T/D$ for a single-spin 2CK
model with explicit left/right lead charge transfer (see
Eq.~\ref{kondo}). Shown for fixed $\rho J_K=10^{-1}$, varying $\rho
J_{LR}=10^{-1},10^{-2}, 10^{-3}, 10^{-4}, 10^{-5}, 10^{-6}$ [lines
(a)--(f)], successively approaching the pure 2CK limit with 
$\rho J_{LR}=0$, line (g). Lower panel: $T=0$ spectra
$D\rho_K(\omega)$ \emph{vs} $\omega/D$ for the same
systems. $S_{\text{imp}}^{1CK}(T)$ and
$\tfrac{1}{2}D\rho_K^{1CK}(\omega)$ for a 1CK
model with $\rho J_K=0.2$ (circles) are also shown for comparison.
}\end{figure}

The above scenario is explored in Fig.~\ref{jlr pert}, where NRG results for Eq.~\ref{kondo} are shown. We
fix $\rho J_K=10^{-1}$, and vary  the  cotunneling term $\rho J_{LR}= 10^{-1}, 10^{-2},10^{-3},10^{-4},10^{-5}$ and
$10^{-6}$ [lines (a)--(f)], approaching the pure 2CK limit $J_{LR}=0$
[line (g)]. The top panel shows the entropy $S_{\text{imp}}(T)/\ln(2)$ \emph{vs} $T/D$, 
from which the behavior associated with the channel-asymmetric
2CK model\cite{2ck:nozieres,2ck:rev_cox_zaw,2ck:BA_sacramento,2ck:BA_andrei_jerez,2ck:CFT_NRG} is seen to arise,
as expected from Eq.~\ref{kondo asym}. In the extreme case $J_K=J_{LR}$ [line (a)], the odd channel is completely decoupled,  no 2CK physics occurs, and the behavior is that of a single-channel Kondo model (circles);
the impurity entropy being completely
quenched below $T\sim T_K^{1CK}$ (cf.\ Eq.~\ref{1CK pms}). 
For smaller $J_{LR}\lesssim T_K^{1CK}$ however, the 2CK scale $T_K^{2CK}$
emerges (cf.\ Eq.~\ref{2CK pms}), below which temperature a 
characteristic $S_{\text{imp}}=\tfrac{1}{2}\ln(2)$ plateau occurs [lines (d)--(g)]. 
Flow to the FL FP below $T\sim T_{\text{FL}}$, for all systems with $J_{LR}\ne 0$, 
is then manifest in the final drop to $S_{\text{imp}}=0$; 
the FL crossover scale $T_{\text{FL}}$ being found to vanish as
\begin{equation}
\label{jlr scale}
T_{\text{FL}}~\overset{J_{LR}\rightarrow 0^{\pm}}
\sim ~ {\cal{A}}|J_{LR}|^{\nu}~~~~~~:\nu =2
\end{equation}
with exponent $\nu =2$, just as expected from mapping to the channel-asymmetric 2CK model, Eq.~\ref{kondo asym}.

We turn now to the $T=0$ spectra $D \rho_K(\omega)\equiv D \rho_{K,\alpha}(\omega)=
-\pi\rho_T\text{Im}[t_{\alpha}(\omega)]$ for lead $\alpha =L$ or $R$
(the two being equivalent by mirror symmetry), shown in the
lower panel of Fig.~\ref{jlr pert}. The spectral behavior is rather
different from what might naively be expected from the entropy, since no low-energy 
$T_{\text{FL}}$ scale is apparent (compare e.g.\ to the channel-asymmetric 2CK models
studied in Fig.~\ref{asym bigj spec}). This however reflects the fact that the model Eq.~\ref{kondo} is channel-asymmetric 2CK in the $e/o$ basis (Eq.~\ref{kondo asym}), rather than the $L/R$ basis; 
and is readily understood using the transformation Eq.~\ref{evenodd}, 
from which one obtains $D\rho_{K,\alpha}(\omega) = 
\tfrac{1}{2}[D\rho_{K,e}(\omega)+D\rho_{K,o}(\omega)]$ in terms of the
spectra for $e/o$ channels. For $J_K=J_{LR}$ [line (a)], 
the odd channel is decoupled in Eq.~\ref{kondo asym}, 
so $D\rho_{K,\alpha}(\omega) = \tfrac{1}{2}D\rho_{K,e}(\omega) \equiv
\tfrac{1}{2} D \rho_K^{1CK}(\omega)$, where $D\rho_K^{1CK}(\omega)$ is the spectrum
for a 1CK model with Kondo coupling $2J_K$ (as confirmed explicitly (circles) in the lower panel of
Fig.~\ref{jlr pert}). 

However, for lines (c)--(f) (corresponding to $J_{LR}\ll T_K^{1CK}$), the spectra are indistinguishable 
from the pure 2CK spectrum, line (g), over the \emph{entire} range of frequencies. 
For $T_{\text{FL}} \ll |\omega| \ll T_K^{2CK}$, RG flow in the vicinity of the 2CK FP 
naturally results in the universal behavior $D\rho_{K,e}(\omega)=D\rho_{K,o}(\omega)=
\tfrac{1}{2}[1-b(|\omega|/T_K^{2CK})^{1/2}]$, as expected for a 2CK
model with small even/odd channel asymmetry (see Fig.~\ref{asym bigj
 spec}).  Consequently $D\rho_{K}(\omega)$ shows the same behavior. For 
$|\omega|\ll T_{\text{FL}}$ by contrast, the spectrum
for the more strongly coupled $e$-channel has the asymptotic form 
$D\rho_{K,e}(\omega)=1-d(|\omega|/T_{\text{FL}})^2$ while the weakly coupled
$o$-channel is described by $D\rho_{K,o}(\omega)=d(|\omega|/T_{\text{FL}})^2$
(see Fig.~\ref{asym bigj spec} and discussion thereof). Indeed, we found in Sec.~\ref{largeJ}
that the \emph{entire} universal crossover to the Fermi liquid FP 
for the strongly coupled lead is related to that of the weakly coupled
lead by $D\rho_{K,e}(\omega)=1-D\rho_{K,o}(\omega)$. Thus, $D\rho_{K}(\omega)=\tfrac{1}{2}$ 
arises for \emph{all} $|\omega|\ll T_K^{2CK}$, as indeed found.
As such, the spectrum $D\rho_{K}(\omega)$ is effectively `blind' to the Fermi
liquid crossover induced by small finite $J_{LR}$: the 2CK FP appears to be stable on the 
lowest energy scales -- although from e.g.\ the entropy we know that this is not the case. 
Ironically, then, experiments that probe the t-matrix (such as measurement of the
zero-bias conductance across dot `1') will always appear to yield 2CK physics,
provided $J_{LR}\lesssim T_K^{1CK}$.


\begin{acknowledgments}
Helpful discussions with M. Galpin and E. Sela are
gratefully acknowledged. This work was in part funded by EPSRC (UK),
under grant EP/D050952/1. AKM also thanks the DFG through SFB 608 and FOR
960 for financial support.
\end{acknowledgments}



\begin{thebibliography}{82}
\bibitem[{not()}]{note:tkdef}
\bibinfo{note}{In practice, $T_K^{SC}$ is determined from a thermodynamic NRG
  calculation via $S_{\text{imp}}(T_K^{SC})=\tfrac{1}{4}\ln(2)$, as such
  halfway between the characteristic 2CK and FL FP values.}
\end{thebibliography}
\end{document}